\documentclass[12pt, a4paper]{article}
\pdfoutput = 1
\usepackage{amsmath,amssymb,bm,epsfig,afterpage}
\usepackage{here}
\usepackage{color}
\usepackage{colortbl}
\usepackage{tabularx}
\usepackage{subcaption}
\usepackage{graphicx}
\usepackage{listings}
\usepackage{longtable}
\usepackage[utf8]{inputenc}

\usepackage[sort&compress,numbers, merge]{natbib}

\definecolor{Orange}{cmyk}{0,0.61,0.87,0}
\definecolor{JungleGreen}{cmyk}{0.99,0,0.52,0}
\definecolor{OliveGreen}{cmyk}{0.64,0,0.95,0.40}
\definecolor{Brown}{cmyk}{0,0.81,1,0.60}
\definecolor{RoyalBlue}{cmyk}{0.71,0.53,0,0.12}
\definecolor{Gray}{cmyk}{0,0,0,0.40}
\definecolor{LightPink}{cmyk}{0.0,0.25,0,0}
\definecolor{LLightPink}{cmyk}{0.0,0.10,0,0}
\definecolor{LightBlue}{cmyk}{0.25,0,0,0}
\definecolor{LightGray}{cmyk}{0,0,0,0.2}

\renewcommand{\thefootnote}{\fnsymbol{footnote}}

\newcommand{\bmat}{\begin{pmatrix} } 
\newcommand{\emat}{\end{pmatrix} }

\allowdisplaybreaks[3]
\newcolumntype{Y}{&gt;{\centering\arraybackslash}X} 

\usepackage[colorlinks=true, linkcolor=OliveGreen, citecolor=RoyalBlue,
urlcolor=RoyalBlue]{hyperref}

\begin{document}

\begin{titlepage}

\begin{flushright}
{\tt 
UT-18-17 \\
EPHOU-18-011
}
\end{flushright}

\vskip 1.35cm
\begin{center}

{\Large
{\bf
Non-Universal Gaugino Masses in the NMSSM
}
}

\vskip 1.5cm

Junichiro~Kawamura$^a$\footnote{
E-mail address: \href{mailto:jkawamur@keio.jp}{\tt jkawamur@keio.jp}},
Tatsuo Kobayashi$^b$\footnote{
E-mail address: \href{mailto:kobayashi@particle.sci.hokudai.ac.jp}{\tt kobayashi@particle.sci.hokudai.ac.jp}}, 
and
Natsumi Nagata$^c$\footnote{
E-mail address: \href{mailto:natsumi@hep-th.phys.s.u-tokyo.ac.jp}{\tt natsumi@hep-th.phys.s.u-tokyo.ac.jp}}

\vskip 0.8cm

{\it $^a$Department of Physics, Keio University, Yokohama
 223--8522, Japan} 
\\[3pt]
{\it 
 $^b$Department of Physics, Hokkaido University, 
Sapporo 060--0810, Japan}
\\[3pt]
{\it $^c$Department of Physics, University of Tokyo, Bunkyo-ku, Tokyo
 133--0033, Japan} 
\date{\today}

\vskip 1.5cm

\begin{abstract}

 The Next-to-Minimal Supersymmetric Standard Model (NMSSM) provides a
 natural framework to realize a low-scale supersymmetric (SUSY) model,
 where a singlet superfield is added to the minimal model to generate a
 SUSY-scale higgsino mass term with its vacuum expectation value. Due to
 the presence of the extra singlet field, the vacuum conditions to
 realize the correct electroweak symmetry-breaking become fairly
 restrictive especially if we impose universality conditions at the unification
 scale. In this paper, we show that a non-universal gaugino mass
 spectrum can significantly relax this restriction even though the
 scalar masses and trilinear couplings are subject to universality
 conditions. With the gaugino non-universality, we find that higgsino
 can be the lightest SUSY particle and its thermal relic
 abundance can reproduce the observed dark matter density in a wide
 range of parameter space in which the 125~GeV Higgs-boson mass is
 obtained. This higgsino-like dark matter may be probed in direct
 detection experiments. We also find that there is an upper bound on the
 masses of supersymmetric particles in this scenario, and many model
 points predict colored particles such as gluino to be within the reach
 of a future 100~TeV collider. Implications for
 no-scale/gaugino-mediation models are also discussed. 

\end{abstract}

\end{center}

\end{titlepage}

\renewcommand{\thefootnote}{\arabic{footnote}}
\setcounter{footnote}{0}

\newcommand{\la}{{\lambda}}
\newcommand{\ka}{{\kappa}}
\newcommand{\mQ}{{m^2_{\tilde{Q}}}}
\newcommand{\mU}{{m^2_{\tilde{u}}}}
\newcommand{\mD}{{m^2_{\tilde{d}}}}
\newcommand{\mL}{{m^2_{\tilde{L}}}}
\newcommand{\mE}{{m^2_{\tilde{e}}}}
\newcommand{\mhu}{{m^2_{H_u}}}
\newcommand{\mhd}{{m^2_{H_d}}}
\newcommand{\ms}{{m^2_S}} 
\newcommand{\Ala}{{A_\lambda}}
\newcommand{\Aka}{{A_\kappa}} 

\section{Introduction} 

A supersymmetric (SUSY) extension of the Standard Model (SM) has been
regarded as the leading candidate for physics beyond the SM. One of the
main reasons for this is its ability to naturally stabilize the weak
scale against radiative corrections if the SUSY breaking scale lies
around the TeV scale. Above this scale, SUSY partners of the SM fields appear and
their contribution to the quantum corrections to the SM Higgs mass
parameter cancels that from the SM particles. This stability is not
spoiled even if extra heavy particles exist at high energies, since the
radiative corrections to the Higgs mass parameter are completely
screened thanks to SUSY. Furthermore, as it turns out, the presence of
the TeV-scale SUSY particles allows the SM gauge coupling constants to
unify at a high-energy scale with their perturbativity maintained up to
the unification scale ($\simeq 2\times 10^{16}$~GeV). Because of these
properties, SUSY theories offer favorable framework to the construction
of a more fundamental theory such as grand unified theories
(GUTs).

In such a high-energy theory, the parameters in the
model are given at the scale of the fundamental theory, say, the
GUT scale. Hence, if the model contains a mass parameter which respects
the symmetries of the theory, we expect this value to be of the order of
the fundamental scale. In the minimal SUSY SM (MSSM), the higgsino mass
parameter, which is called the $\mu$ parameter, has such a property
\cite{Kim:1983dt}. It turns out however that if the $\mu$ parameter is
much larger than the SUSY-breaking scale the electroweak symmetry
breaking (EWSB) does not occur and thus the model cannot be considered
to be realistic. A simple way to evade this problem is to remove this
mass parameter from the theory with the help of an extra symmetry. We
then add a singlet superfield and couple this to the Higgs superfields
so that it generates an effective $\mu$ parameter after it develops a
vacuum expectation value (VEV). This setup is dubbed as the
Next-to-Minimal SUSY SM (NMSSM) \cite{Ellwanger:2009dp, Maniatis:2009re}. 

In the NMSSM, both the singlet and the MSSM Higgs fields can acquire VEVs
only after SUSY is broken. This can thus explain why these values are as
small as the weak scale if the SUSY-breaking scale is around the TeV
scale. In fact, it is known \cite{Ellwanger:2009dp, Maniatis:2009re}
that with an appropriate choice of SUSY parameters we can obtain a
desirable vacuum at which the electroweak gauge symmetry is
adequately broken so that the observed mass spectrum in the SM is
realized. This observation makes the NMSSM a quite promising
candidate for the SUSY SM. 

This vacuum condition, however, turns out to be highly restrictive once
we consider some universality at the input scale. In the case of the
MSSM, it has been widely known that even though we assume universality among
the parameters at the input scale, as in the Constrained Minimal SUSY SM
(CMSSM), we can easily obtain a viable set of model parameters at low
energies (for recent studies of such models, see, {\it e.g.},
Refs.~\cite{Ellis:2012aa, Baer:2012uya,  Baer:2012mv, Ellis:2012nv,
Liu:2013ula, Buchmueller:2013psa, Buchmueller:2013rsa,
Roszkowski:2014wqa, Buchmueller:2015uqa, Bagnaschi:2015eha,
Ellis:2015rya, Athron:2017qdc, Ajaib:2017iyl, Costa:2017gup} and
references therein). The NMSSM counterpart of the CMSSM, called CNMSSM,
has also been extensively studied so far \cite{Ellis:1988er, Ellwanger:1993xa,
Elliott:1994ht, Ellwanger:1995ru, King:1995vk, Ellwanger:1996gw,
Djouadi:2008yj, Djouadi:2008uj, Arbey:2011ab}. 
In this case, the correct
electroweak vacuum can be realized only in the limited 
parameter space, where the universal scalar mass $m_0$ and trilinear
parameter $A_0$ satisfy $ 3m_0 \lesssim |A_0|$ and both
are much smaller than the universal gaugino mass $m_{1/2}$. These
relations make it rather difficult to obtain the observed value of the
mass of the SM-like Higgs boson~\cite{Arbey:2011ab}, $m_h \simeq
125$~GeV~\cite{Aad:2015zhl}, and require SUSY-breaking scale to be much
higher than the electroweak scale. In 
addition, the condition $ 3m_0 \lesssim |A_0|$ tends to make stau
the lightest SUSY particle (LSP) or even tachyonic. In the region where
this is evaded, the LSP is singlino---though in principle this can be a
good dark matter candidate, its thermal relic abundance often exceeds the
observed value of dark matter abundance, $\Omega_{\text{DM}} h^2 \simeq
0.12$~\cite{Aghanim:2018eyx}, especially when SUSY-breaking scale is
in the multi-TeV range so that the 125~GeV Higgs-boson mass is
obtained. Such over abundance requires non-trivial cosmological
history like entropy production by a long-lived particle. To evade these
problems, it is often the case that the universality conditions on the
singlet scalar masses and/or trilinear terms are relaxed so that the correct
electroweak vacuum is easily obtained \cite{Ellwanger:2006rn, Gunion:2012zd,
Ellwanger:2012ke, Kowalska:2012gs, Das:2013ta, Beskidt:2013gia,
Ellwanger:2014dfa, Beskidt:2016egy}. The introduction of right-handed
neutrinos may also improve the situation through renormalization group (RG)
effect due to their couplings to the singlet superfield
\cite{Cerdeno:2017sks}. 

In this paper, we discuss another possibility, namely, a
constrained NMSSM with non-universal gaugino masses. A non-universal
gaugino mass spectrum can be realized in theoretically well-motivated
frameworks such as the mirage mediation \cite{Choi:2004sx, Choi:2005ge,
Endo:2005uy, Choi:2005uz, Choi:2005hd, Kitano:2005wc, Choi:2006xb},
non-minimal GUT models \cite{Ellis:1984bm, Ellis:1985jn, Drees:1985bx, Anderson:1996bg, Chakrabortty:2008zk, Martin:2009ad, Chakrabortty:2010xq, Younkin:2012ui, Kobayashi:2017fgl}, models with
a non-universal gauge kinetic function via string compactifications
\cite{Blumenhagen:2006ci, Ibanez:2012zz}, and so on, and found to be advantageous for
the electroweak naturalness problem \cite{Abe:2007kf, Abe:2012xm}. These
observations have stimulated many studies on non-universal gaugino masses so far
\cite{Bhattacharya:2007dr, Bhattacharya:2009wv, Horton:2009ed, Brummer:2012zc, Gogoladze:2012yf,
Yanagida:2013ah, Gogoladze:2013wva, Yanagida:2013uka, Chakrabortty:2013voa, Martin:2013aha,
Chakrabortty:2015ika, Harigaya:2015jba, Abe:2015xva, Sumita:2015tba,
Kawamura:2016drh, Kawamura:2017amp, Kawamura:2017qey,
Martin:2017vlf}. In particular, discussions on mirage mediation in
the NMSSM can be found in Refs.~\cite{Kobayashi:2012ee, Asano:2012sv,
Hagimoto:2015tua}. The present paper focuses on the implications of
non-universal gaugino mass spectrum for the vacuum condition in the
NMSSM with universal conditions on scalar masses and trilinear couplings
imposed at the unification scale. We find 
that gaugino non-universality modifies the vacuum
condition through RG effects and significantly
enlarges the parameter space where the desirable EWSB
is realized. With this modification, higgsino turns
out to be the LSP and its thermal relic abundance can account for the
observed dark matter density in a wide range of the parameter
space. This higgsino-like dark matter may be probed in dark matter
direct detection experiments. Moreover, by requiring the thermal relic
of the LSP to be equal to or smaller than $\Omega_{\text{DM}} h^2$, we
obtain an upper bound on the mass scale of SUSY particles, and many of
the viable parameter points are found to predict colored particles such
as gluino to be within the reach of a future 100~TeV
collider \cite{Cohen:2013xda, Ellis:2015xba, Arkani-Hamed:2015vfh,
Golling:2016gvc}. 

This paper is organized as follows. In the next section, we give a brief
review on the vacuum conditions in the NMSSM, and discuss the effect of a
non-universal gaugino mass spectrum on these conditions. Then, we
study the phenomenological implications of this setup in
Sec.~\ref{sec:pheno}, followed by an analysis devoted to
no-scale/gaugino-mediation like spectra in
Sec.~\ref{sec:noscale}. Section~\ref{sec:conclusion} is for conclusion
and discussions.

\section{Vacuum conditions in NMSSM}
\label{sec:conds}

\subsection{Scalar sector in NMSSM}
\label{sec:vaccond}

To begin with, we review the scalar sector in the NMSSM with a
particular focus on the vacuum conditions. For previous studies on the
scalar sector in the NMSSM, see Refs.~\cite{Pandita:1993tg,
Elliott:1993bs, Ellwanger:1996gw, Ellwanger:1999bv, Miller:2003ay,
Funakubo:2004ka, Cheung:2010ba, Kanehata:2011ei, Kobayashi:2012xv, Beuria:2016cdk}.
Throughout this paper we consider the so-called $\mathbb{Z}_3$-invariant
NMSSM, which is characterized by the superpotential of the Higgs sector 
\begin{equation}
 W_{\rm Higgs} = 
\lambda S (H_u \cdot H_d) + \frac{1}{3}\kappa S^3 ~,
\end{equation}
and the corresponding soft terms
\begin{align}
 {\cal L}_{\rm soft}^{(\rm Higgs)} = &-m_{H_u}^2 |H_u|^2 -m_{H_d}^2 |H_d|^2 - m_S^2
 |S|^2 
- \biggl[
\lambda A_\lambda S(H_u \cdot H_d) + \frac{1}{3} \kappa A_\kappa S^3 
+{\rm h.c.}
\biggr]~.
\end{align}
As the name stands for, this theory possesses an exact 
$\mathbb{Z}_3$ symmetry at the Lagrangian level, which is spontaneously
broken when the scalar fields develop VEVs. The spontaneous breaking
of a discrete symmetry results in the generation of domain walls
\cite{Zeldovich:1974uw, Vilenkin:1984ib, Ellis:1986mq, Abel:1995wk}, which are
cosmologically harmful. To evade this problem, one often introduces
Planck-scale suppressed non-renormalizable operators that explicitly
break the $\mathbb{Z}_3$ symmetry in a proper way that these operators
do not induce sizable tadpole contributions at low energies
\cite{Abel:1996cr, Panagiotakopoulos:1998yw, Panagiotakopoulos:1999ah,
Panagiotakopoulos:2000wp, Dedes:2000jp}. In the presence of such an
explicit $\mathbb{Z}_3$-breaking term, domain walls become
unstable.\footnote{For recent studies on unstable domain walls in the
NMSSM, see Refs.~\cite{Hamaguchi:2011nm, Kadota:2015dza,
Hattori:2015xla, Saikawa:2017hiv}.}
In the following discussion, we
implicitly assume such a mechanism, which has no effect on our argument
presented in this paper.
The scalar potential for the neutral fields is obtained from these
Lagrangian terms as 
\begin{align}
 V_{\rm neutral} &
=
m_S^2 |S|^2
+ 
\left(|\lambda S|^2 + m_{H_u}^2\right)  |H_u^0|^2 + 
\left(|\lambda S|^2 + m_{H_d}^2\right) 
|H_d^0|^2 +  \left|  \kappa S^2- \lambda H_u^0 H_d^0  \right|^2
 \nonumber \\[3pt]
&+ \frac{1}{8} (g^2 +g^{\prime 2}) \left(
|H_u^0|^2 - |H_d^0|^2
\right)^2
+\biggl[
 - \lambda A_\lambda SH_u^0 H_d^0 + \frac{1}{3} \kappa A_\kappa S^3 
+{\rm h.c.}
\biggr]~.
\label{eq:scapot}
\end{align}
In what follows, we take all of the parameters to be real just for
simplicity. From this potential, we readily find the tadpole
conditions for the scalar fields $H_u$, $H_d$, and $S$:
\begin{align}
 & \frac{1}{2}\lambda^2 v_u (v_d^2 + v_s^2) 
+ \frac{1}{8} v_u (g^2 + g^{\prime 2}) (v_u^2 -v_d^2)
+ m_{H_u}^2 v_u  -\frac{1}{2}\lambda \kappa v_d v_s^2 
-\frac{1}{\sqrt{2}}\lambda A_\lambda v_d v_s    = 0 ~,
\label{eq:tadhuz3}\\[3pt]
 & \frac{1}{2}\lambda^2 v_d (v_u^2 + v_s^2) 
+ \frac{1}{8} v_d (g^2 + g^{\prime 2}) (v_d^2 -v_u^2)
+ m_{H_d}^2 v_d  -\frac{1}{2}\lambda \kappa v_u v_s^2 
-\frac{1}{\sqrt{2}}\lambda A_\lambda v_u v_s   = 0 ~,
\label{eq:tadhdz3}\\[3pt]
&  m_S^2 v_s
+ \kappa^2 v_s^3 
+ \frac{1}{\sqrt{2}}\kappa A_\kappa v_s^2 
-\frac{1}{\sqrt{2}} \lambda A_\lambda v_u v_d 
- \lambda \kappa  v_u v_d v_s 
+ \frac{1}{2}  \lambda^2 v^2 v_s 
 =0~, 
\label{eq:tadhsz3}
\end{align}
where $\langle H_u \rangle = v_u/\sqrt{2}$, $\langle H_d \rangle =
v_d/\sqrt{2}$, $\langle S \rangle = v_s/\sqrt{2}$, and $v^2 \equiv v_u^2
+ v_d^2 \simeq 246$~GeV. A non-zero VEV of $\langle S \rangle$ gives an
effective higgsino mass term, \textit{i.e.}, $\mu$-parameter: $
\mu_{\text{eff}} \equiv \lambda \langle S \rangle = 
\lambda v_s/\sqrt{2}$. 
As usual, we take $v_u$, $v_d$, and $\lambda$ to be positive without
loss of generality. The sign of $\mu_{\text{eff}}$ follows that of
$v_s$, which can be either positive or negative. 
For later use, we further rewrite Eqs.~\eqref{eq:tadhuz3} and \eqref{eq:tadhdz3}
in the following form:
\begin{align}
&\mu_{\text{eff}}^2 
+\frac{1}{2} m_Z^2 + \frac{m_{H_u}^2 \tan^2 \beta -m_{H_d}^2 }{\tan^2 \beta-1}
=0 ~,
 \label{eq:vaccond1}
\\[3pt]
& \left[
m_{H_u}^2 + m_{H_d}^2 + 2\mu_{\text{eff}}^2 +\frac{1}{2}\lambda^2 v^2 
\right]\sin 2\beta
-2\mu_{\text{eff}}B_{\text{eff}} = 0 ~,
 \label{eq:vaccond2}
\end{align}
where $\tan\beta \equiv v_u/v_d$, $m_Z = v\sqrt{g^2 + g^{\prime 2}}/2$,
and $B_{\text{eff}} \equiv A_\lambda + \kappa v_s/\sqrt{2}$.

To see the condition for $v_s \neq 0$, let us study the tadpole
condition \eqref{eq:tadhsz3} in the limit of $v_s \gg v_u, v_d$ to keep
the first three terms:
\begin{equation}
 \left[
m_S^2
+ \kappa^2 v_s^2 
+ \frac{1}{\sqrt{2}}\kappa A_\kappa v_s
\right]v_s \simeq 0 ~.
\end{equation}
This has non-zero real solutions if and only if $A_\kappa^2 \gtrsim 8
m_S^2$, which are given by
\begin{equation}
  v_s^{(\pm)} \simeq \frac{1}{2\sqrt{2} \kappa} 
\left[-A_\kappa \pm \sqrt{A_\kappa^2 - 8 m_S^2}\right] ~.
\label{eq:vssol}
\end{equation}
At either of them, the scalar potential should be deeper than at the
origin so that $v_s \neq 0$ is energetically favored. For a negative
(positive) $A_\kappa$, the potential value at $v_s^{(+)}$ is smaller
(larger) than that at $v_s^{(-)}$, and in both of these cases it is
below that at the origin if
\begin{equation}
 A_\kappa^2 \gtrsim 9\ m_S^2 ~.
\label{eq:condakms}
\end{equation}
This condition is trivially satisfied if $m_S^2 < 0$. If, on the other
hand, $m_S^2 > 0$, then a rather large $A_\kappa$ is required. 

For $v_s^{(\pm)}$ to be a minimum (not a saddle point), 
the second Hessian matrix of the scalar potential with respect to the scalar fields should be positive definite. 
As long as the singlet-doublet
mixing is not so large, the curvatures in the $H_u$ and $H_d$ directions are
similar to those in the MSSM, and thus only the singlet direction is
potentially dangerous. The curvature in this direction is read from the
masses of the singlet scalar and pseudo-scalar, which are respectively
given by 
\begin{align}
 m_s^2 &\simeq 2 \kappa^2 v_s^2 + \frac{\kappa A_\kappa}{\sqrt{2}}  v_s ~,
\\
 m_a^2 &\simeq - \frac{3\kappa A_\kappa}{\sqrt{2}} v_s ~.
\label{eq:ma2}
\end{align}
These masses should be positive, which yield additional constraints. 

In the NMSSM, due to the complexity of the scalar potential, there are
quite a few potential minima other than the desired one we considered above. In
this paper, we require this desired minimum to be the global minimum, though this is not necessary as long as the lifetime of
the vacuum is sufficiently larger than the age of the Universe.
For a recent study on metastable vacua in the NMSSM, see
Ref.~\cite{Beuria:2016cdk}. Possibilities of unwanted minima in the
NMSSM are considered in Refs.~\cite{Kanehata:2011ei, Kobayashi:2012xv,
Beuria:2016cdk}. Here, we summarize some important cases among
them:\footnote{As can be seen from Eqs.~\eqref{eq:tadhuz3},
\eqref{eq:tadhdz3}, and \eqref{eq:tadhsz3}, if one of the scalar fields
has the zero VEV, then at least one of the other two scalar fields must
also have the vanishing VEV unless there are unrealistic particular
relations among the parameters in the scalar potential---namely, it is
generically not possible that only two among the three scalar fields
acquire VEVs.  } 
\begin{itemize}
 \item[(a)] A potentially dangerous minimum may be found along the
	    direction in which both the $F$- and $D$-terms vanish, which
	    occurs when
\begin{equation}
 |H_u^0| = |H_d^0|~, \qquad \kappa S^2 = \lambda H_u^0 H_d^0 ~. 
\label{eq:fdflat}
\end{equation}
	    In this case, the fourth and fifth terms in Eq.~\eqref{eq:scapot}
	    vanish. By taking $\lambda$, $H_u^0$, and $H_d^0$ to be real
	    and positive without loss of generality, we have the scalar
	    potential in the form 
\begin{align}
 V_{F,D} (\phi) = \left(m_{H_u}^2 + m_{H_d}^2 + \frac{\lambda}{\kappa}
 m_{S}^2\right) \phi^2 
\pm 2\lambda \sqrt{\frac{\lambda}{\kappa}} \left[
-A_\lambda + \frac{1}{3} A_\kappa
\right] \phi^3 + \frac{2\lambda^3}{\kappa} \phi^4 
~,
\end{align}
	    where we set $\phi \equiv H_u^0 = H_d^0$. Notice that
	    $\kappa \geq 0$ follows from the second condition in
	    Eq.~\eqref{eq:fdflat} with this convention. The field $\phi$
	    can have a non-trivial minimum if 
\begin{equation}
 9 \left[
-A_\lambda + \frac{1}{3} A_\kappa
\right]^2 \geq 16\left(m_{H_u}^2 + m_{H_d}^2 + \frac{\lambda}{\kappa}
 m_{S}^2\right) ~.
\label{eq:dfflatlocalcond}
\end{equation}
	    If this holds, then we need to make sure that the height of
	    $V_{F,D}$ at the minimum be higher than the potential value
	    at the desired minimum. Generically speaking, if soft
	    masses, especially $m_{H_d}^2$, are large enough compared to
	    the $A$-terms $A_\lambda^2$ and $A_\kappa^2$, this direction
	    is stabilized. 

 \item[(b)] Another direction which may provide a local minimum is along
	    the $D$-flat direction but away from the $F$-flat
	    direction. As shown in Ref.~\cite{Kobayashi:2012xv},  we can
	    obtain a condition 
	    similar to Eq.~\eqref{eq:dfflatlocalcond} and
	    this direction is again stabilized if scalar masses are
	    sufficiently larger than the $A$-terms. This direction
	    includes a special case where $H_u^0 = H_d^0 = 0$ and $S\neq
	    0$.

 \item[(c)] We may also consider the case where the $F$-term vanishes
	    but the $D$-term has a non-zero value. Again, there is a
	    condition similar to Eq.~\eqref{eq:dfflatlocalcond} for the
	    presence of a local minimum in this direction
	    \cite{Kobayashi:2012xv}, and the minimum disappears if
	    scalar masses are larger than the $A$-terms squared. This
	    direction contains two special cases where 
	    $S = H_d^0 = 0$ and $H_u^0\neq 0$, or $S = H_u^0 = 0$ and
	    $H_d^0\neq 0$. In the former case, the scalar potential has
	    the form 
\begin{equation}
 V_{S,D}(H_u^0) = m_{H_u}^2 |H_u^0|^2 + \frac{g^2 +g^{\prime 2}}{8}
  |H_u^0|^4 ~.
\label{eq:vsd}
\end{equation}
	    This potential has a non-trivial solution if $m_{H_u}^2 <
	    0$ and the potential value at this minimum is given by
\begin{equation}
 V_{S,D}^{(\text{min})} = -\frac{2 (m_{H_u}^2)^2}{g^2 + g^{\prime 2}} ~.
\label{eq:vsdmin}
\end{equation}
	    As we see, if $|m_{H_u}^2|$ is very large, then this may become the
	    global minimum. In the latter case, on the other hand, we
	    obtain a similar expression to Eq.~\eqref{eq:vsd} and find
	    that there is no minimum in this direction since $m_{H_d}^2$
	    is always positive in the parameter space we are interested
	    in. 

\item[(d)] We also need to ensure the absence of charge and/or color
	   breaking minima, at which some sfermions acquire non-zero
	   VEVs. To avoid such a minimum, we require
	   \cite{Derendinger:1983bz, Stephan:1997ds, Ellwanger:1999bv}
\begin{align}
 A_{u_i}^2 & \leq 3 \bigl(m_{H_u}^2 + m_{\widetilde{Q}_i}^2 +
 m_{\widetilde{D}_i}^2 \bigr) \quad \text{at scale}~~\mu \sim A_{u_i} /
 y_{u_i} ~,\\[2pt]  
 A_{d_i}^2 & \leq 3 \bigl(m_{H_d}^2 + m_{\widetilde{Q}_i}^2 +
 m_{\widetilde{U}_i}^2 \bigr) \quad \text{at scale}~~\mu \sim A_{d_i} /
 y_{d_i} ~,\\[2pt]  
 A_{e_i}^2 & \leq 3 \bigl(m_{H_d}^2 + m_{\widetilde{L}_i}^2 +
 m_{\widetilde{E}_i}^2 \bigr) \quad \text{at scale}~~\mu \sim A_{e_i} /
 y_{e_i} ~,
\end{align}
where $A_f$ ($f = u_i, d_i, e_i$) are the trilinear couplings corresponding
to the SM Yukawa couplings $y_f$ and $m_{\tilde{f}}$ stand for
the sfermion masses. Again, we see that these conditions are satisfied
	   as long as the
	   $A$-terms are much smaller than the soft masses.

\end{itemize}

\subsection{CNMSSM}
\label{sec:CMSSM}

Next, we give a
brief review on the
CNMSSM, which is stringently constrained by the vacuum conditions. In
the CNMSSM, the gaugino masses,  soft scalar masses, and $A$-terms are
respectively taken to be universal at the GUT scale $M_{\text{GUT}}$,
which is defined by the condition $g_1 (M_{\text{GUT}}) = g_2
(M_{\text{GUT}})$:  
\begin{align}
 M_1 &= M_2 = M_3 \equiv m_{1/2} ~, \\[3pt]
 m_{\tilde{f}}^2 &= m_{H_u}^2 = m_{H_d}^2 = m_S^2 \equiv m_0^2 ~,\\[3pt]
 A_f &= A_\lambda = A_\kappa \equiv A_0 ~,
\label{eq:univcond}
\end{align}
where $M_a$ ($a=1,2,3$) are gaugino masses. The absolute value of the singlet VEV
$|v_s|$, the singlet trilinear coupling $\kappa$, and $\tan\beta$ are
determined such that the tadpole conditions \eqref{eq:tadhuz3},
\eqref{eq:tadhdz3}, and \eqref{eq:tadhsz3} are satisfied with the
correct size of the Higgs VEV $v \simeq 246$~GeV. Note that the first
and second parameters correspond to the degrees of freedom of $|\mu|$
and $B\mu$ in the case of the CMSSM. The third parameter, $\tan\beta$,
is fixed by the universality condition $m_S^2 = m_0^2$ at the GUT
scale. As a result, the free input parameters in the CNMSSM are 
\begin{equation}
 m_0, \quad m_{1/2}, \quad A_0,  \quad \lambda, \quad 
 \text{sign}(v_s) ~.
\label{eq:inppar}
\end{equation}
Thus, the number of free parameters in the CNMSSM is the same as in the
CMSSM, where each point in the parameter space is specified by $m_0$,
$m_{1/2}$, $A_0$, $\tan \beta$, and $\text{sign}(\mu)$. 

It turns out that with the universal conditions \eqref{eq:univcond} it
is rather difficult to assure the vacuum conditions given in the
previous subsection to be satisfied. To evade the current LHC limits on the
masses of SUSY particles as well as to explain the observed Higgs
boson mass, generically speaking, we need to take the soft SUSY-breaking
parameters in Eq.~\eqref{eq:inppar} to be ${\cal O}(1)$~TeV or
larger. Then, for a moderate or large value of $\tan \beta$,
Eq.~\eqref{eq:vaccond1} is approximated by
\begin{equation}
 \mu_{\text{eff}}^2 = \frac{1}{2} \lambda^2 v_s^2 \simeq - m_{H_u}^2 ~.
  \label{eq:simpvaccond1}
\end{equation}
For $m_{1/2} \gtrsim {\cal O}(1)$~TeV, $ m_{H_u}^2$ is driven to be a
large negative value through the RG effect by gluino
so that $- m_{H_u}^2 \simeq m_{1/2}^2$. Thus, $\mu_{\text{eff}}^2 \simeq
m_{1/2}^2$ is required from the vacuum condition. If both $\lambda$ and
$\kappa$ are sizable, this condition and Eq.~\eqref{eq:vssol} imply that
a large value of $|A_\kappa|$ or $- m_S^2$ is required so that $|v_s| $
can be as large as ${\cal O}(m_{1/2})$. We however notice that it is
difficult to obtain a large $|A_\kappa|$ at low energies as it is
suppressed via the RG effect in the case where $\lambda$
and $\kappa$ are sizable, while it is hard to reconcile a large negative
value of $m_S^2$ at low energies with the universality condition $m_S^2
= m_0^2 > 0$ at the input scale. To see this in more qualitative manner,
we show the soft mass parameters at a low energy scale that are relevant
to the vacuum conditions as functions of the parameters at the GUT
scale:
\begin{align}
 A_\lambda (M_s) = &-0.019 M_1-0.252 M_2 +0.450 M_3 +0.231 A_0 ~,
 \label{eq-AlB} \\[3pt]  
 A_\kappa (M_s) = &~ 0.002 M_1+0.015 M_2-0.012 M_3 + 0.408 A_0 ~,
 \label{eq-AkB} \\[3pt] 
 m_S^2 (M_s) = & -0.001M_1^2 -0.010M_2^2 +
                        0.003M_2M_3+0.010M_3^2 \nonumber \\ 
                    &+A_0 (0.002 M_2-0.006M_3) 
                        -0.076A_0^2 +0.410m_0^2 ~,\label{eq-msB}\\[3pt]
 \mhu(M_s)=&~ 0.010 M_1^2 -0.004 M_1M_2+0.210M_2^2-0.013 M_1M_3 
                       -0.078 M_2M_3 -0.902 M_3^2  \nonumber \\  
                      &+
 A_0(0.010M_1+0.054M_2+0.193M_3)-0.104A_0^2+0.095m_0^2 ~,\label{eq-mhuB} \\[3pt] 
\mhd(M_s)=&~ 0.014 M_1^2-0.004M_1M_2+0.230M_2^2-0.003M_1M_3-0.062M_2M_3
                       -0.603M_3^2  \nonumber \\ 
                    &
 +A_0(0.011M_1+0.062M_2+0.159M_3)-0.123A_0^2+0.192m_0^2 ~,\label{eq-mhdB} 
\end{align}
where we set $\lambda = 0.2$, $\kappa = 0.5$, $\tan \beta =50$, and $M_s
= 7.5$~TeV. We have evaluated the numerical coefficients of the above
expressions by solving the two-loop renormalization group equations
(RGEs) without imposing the universality conditions in the CNMSSM. 
It is found from Eq.~\eqref{eq-mhuB} that the low-energy value
of $m_{H_u}^2$ is basically determined by the gluino mass and thus
becomes a large negative value when $M_3$ is large. On the other hand,
both $A_\kappa$ and $m_S^2$ at low energies tend to be relatively suppressed as
seen in Eq.~\eqref{eq-AkB} and Eq.~\eqref{eq-msB}, respectively, with
which $\mu_{\text{eff}}^2 \ll - m_{H_u}^2$ and thus the vacuum condition
\eqref{eq:vaccond1} cannot be satisfied. 

One might think that a sufficiently large $\mu_{\text{eff}}^2$  can be
obtained where $|\kappa| \ll \lambda$, since $|v_s|$ becomes large when
$|\kappa|$ is small, as can be seen from
Eq.~\eqref{eq:vssol}. Such a parameter region is, however, disfavored by
the vacuum stability condition. To see this, let us compare the height of
the potential at a desired vacuum $v_u, v_d, v_s \neq 0$ with $V_{S,
D}^{(\text{min})}$ given in Eq.~\eqref{eq:vsdmin}. In the limit of $v_s
\gg v_u, v_d$, the former is computed as 
\begin{align}
 V_{\text{min}} &\simeq \frac{1}{2} m_S^2 v_s^2 + \frac{1}{3\sqrt{2}}
 \kappa A_\kappa v_s^3 + \frac{1}{4} \kappa^2 v_s^4 \nonumber \\
& = -\frac{\kappa^2}{4} v_s^4 - \frac{1}{6\sqrt{2}}\kappa A_\kappa v_s^3 ~,
\end{align}
where we have eliminated $m_S^2$ in the last equation using
Eq.~\eqref{eq:vssol}. On the other hand, with the
condition \eqref{eq:simpvaccond1} $V_{S, D}^{(\text{min})}$ can
be approximated by 
\begin{equation}
 V_{S, D}^{(\text{min})} \simeq - \frac{\lambda^4 v_s^4}{2(g^2 +
  g^{\prime 2})} ~.
\end{equation}
Thus, $V_{\text{min}} < V_{S, D}^{(\text{min})}$ requires
\begin{equation}
 \frac{\lambda^4}{g^2 +  g^{\prime 2}} < \frac{\kappa^2}{2} + 
\frac{\kappa A_\kappa}{3 \sqrt{2} v_s} ~,
\end{equation}
which cannot be satisfied if $|\kappa| \ll \lambda$. 

The above difficulties can be avoided in the parameter space where both
$\lambda$ and $|\kappa|$ are very small. In this case, both $A_\kappa$
and $m_S^2$ rarely run in the RG flow, and thus $A_\kappa \simeq A_0$
and $m_S \simeq m_0$. The condition \eqref{eq:condakms} then leads to
$|A_0| \gtrsim 3 m_0$, which assures the existence of a solution for
$v_s \neq 0$. Even though $\lambda \ll 1$, the condition \eqref{eq:simpvaccond1}
may be satisfied by taking $|\kappa|$ to be also very small so that
the suppression in $|\mu_{\text{eff}}|$ by a small $\lambda$ is
compensated by an enhancement in $|v_s|$ with $|\kappa|<\lambda \ll 1$. 
In fact, a detailed
study performed in Ref.~\cite{Djouadi:2008uj} demonstrates that viable
model points are found in the parameter region where (a) $|\kappa| <
\lambda \ll 1$; (b) $3m_0 \lesssim |A_0|$; (c) $m_{1/2} \gg m_0$; (d) 
$\tan\beta \gg 1$. The reason for the first two conditions have already
been addressed. The condition (c) is required in order to avoid the stau
LSP or the presence of a tachyonic particle, which often occurs when
$|A_0|/m_0$ is sizable. When the conditions (a--c) are met, the second term in
Eq.~\eqref{eq:vaccond2} tends to be much smaller than the sum of the
terms in the square brackets of the first term, which then leads to
$\tan \beta \gg 1$. 

In the phenomenologically viable region found in
Ref.~\cite{Djouadi:2008uj}, the LSP is a singlino-like neutralino. In
principle, this can be a good dark matter candidate, but in practice
this may cause a problem as its thermal relic abundance tends to be much
larger than the observed dark matter density due to its small
annihilation cross section. As discussed in Ref.~\cite{Djouadi:2008uj},
the correct dark matter density is obtained only in the limited region
of the parameter space where the lighter stau is degenerate with the
singlino LSP in mass so that the dark matter abundance is efficiently
reduced via coannihilation \cite{Griest:1990kh}. It turns out that the
correct relic abundance can be obtained if $m_{1/2} \lesssim \text{a few
TeV}$. With such a small $m_{1/2}$, however, it in turn becomes difficult to
explain the observed value of the Higgs boson mass, $m_h \simeq
125$~GeV. Indeed, a recent analysis \cite{Cerdeno:2017sks} shows that
the parameter points consistent with $m_h \simeq 125$~GeV are obtained
only for $m_{1/2} \gtrsim 3.5$~TeV, where the stau-coannihilation is no longer
sufficiently effective. This result implies that we need some
additional mechanism to reduce the dark matter abundance, such as the
late-time entropy production.

\subsection{Effect of non-universal gaugino masses}

Now we discuss the effect of non-universal gaugino masses on the
vacuum conditions presented in Sec.~\ref{sec:vaccond} and compare this
result with that of the CNMSSM discussed in the previous
subsection. As we see above, an obstacle to the vacuum conditions in the
CNMSSM is a large negative value of $m_{H_u}^2$ due to the RG effect by
gluino. However, this contribution can be canceled by the wino
contribution once we allow non-universal gaugino masses
\cite{Abe:2007kf, Abe:2012xm}. This feature can easily be seen by
examining Eq.~\eqref{eq-mhuB}; if we take $M_2 = \text{a few} \times
M_3$, the gaugino contributions cancel with each other so that
$|m_{H_u}^2|$ is much smaller than that in the case of the universal
gaugino masses.  Moreover, Eq.~\eqref{eq-msB} shows that a heavy wino
gives a negative contribution to $m_S^2$, which again relaxes the
constraint from the vacuum conditions with the universality condition
$m_S^2 = m_0^2$ at $M_{\text{GUT}}$---the negative wino contribution can
make $m_S^2$ run negative at low energies even though it is positive at
$M_{\text{GUT}}$, which allows the radiative generation of a non-zero
$v_s$. 

\begin{figure}
\centering
\includegraphics[height=75mm]{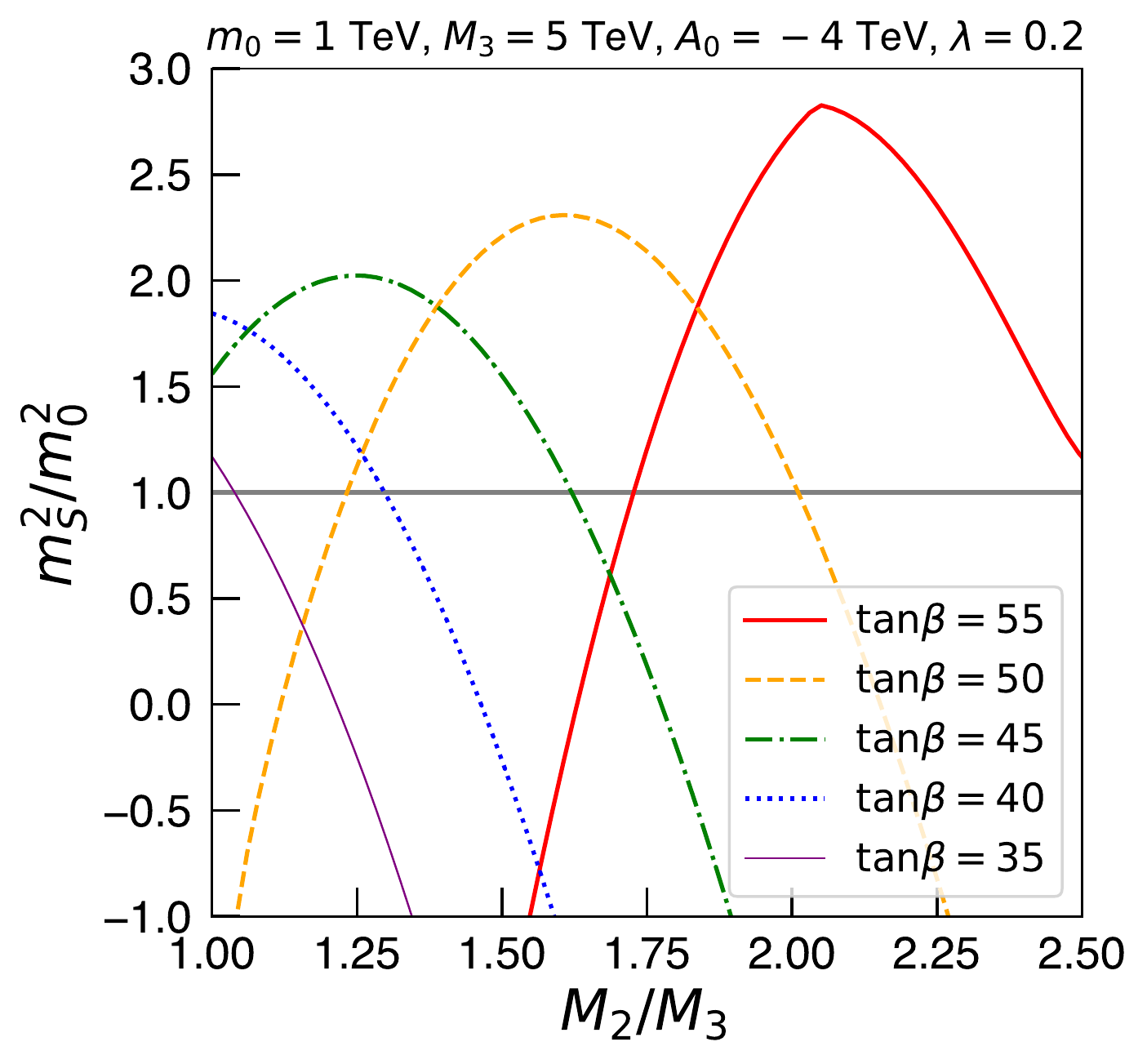}
\caption{$m_S^2/m_0^2$ at the GUT scale as a function of the gaugino
 mass ratio $M_2/M_3$ for different values of $\tan\beta$. Here, we take
 $m_0 = 1$~TeV, $M_3 = 5$~TeV, $A_0= -4$~TeV, and $\lambda =0.2$. 
}  
\label{fig:r23vsms}
\end{figure}

To see the effect of non-universal gaugino masses on the universality
condition $m_S^2 = m_0^2$ at $M_{\text{GUT}}$ in more detail, in
Fig.~\ref{fig:r23vsms}, we show $m_S^2/m_0^2$ at the GUT scale as a
function of the gaugino mass ratio $M_2/M_3$ for different values of
$\tan\beta$. Here, we take $m_0 = 1$~TeV, $M_3 = 5$~TeV, $A_0= -4$~TeV,
and $\lambda =0.2$. To obtain the GUT-scale value of $m_S^2$, we first
determine its SUSY-scale value using the vacuum conditions, and evolve
it up to $M_{\text{GUT}}$ according to RGEs; we exploit
\texttt{NMSSMTools 5.3.1} \cite{Ellwanger:2004xm, Ellwanger:2005dv,
Ellwanger:2006rn} for this purpose. We here regard $\tan\beta$ as a free
parameter by relaxing the universality condition $m_S^2 = m_0^2$. 
The other gaugino mass ratio, $M_1/M_3$, is fixed such that the
gaugino masses satisfy the following relation, which is motivated by the 
mixed modulus-anomaly mediation \cite{Choi:2005uz}:\footnote{From this
relation, we have
\begin{equation}
 \frac{M_1}{M_3} = \frac{1}{b_2 -b_3} \biggl[
(b_1-b_3) \frac{M_2}{M_3} + b_2 -b_1
\biggr]~.
\end{equation}
}
\begin{equation}
 M_a (M_{\text{GUT}}) = M_0 \biggl[1 + \frac{\alpha b_a
  g_{\text{GUT}}^2}{16\pi^2} \ln
  \biggl(\frac{M_{\text{Pl}}}{m_{3/2}}\biggr)\biggr] ~,
\label{eq:gauginomasses}
\end{equation}
where 
$(b_1, b_2, b_3) = (33/5, 1, -3)$ are the one-loop gauge-coupling
beta-function coefficients,
$g_{\text{GUT}}$ denotes the unified gauge coupling,
$M_{\text{Pl}}$ is the reduced Planck mass,
$m_{3/2}$ is the gravitino mass,
$M_0$ denotes the modulus-mediated contribution to the gaugino mass, and
$\alpha$ is an ${\cal O}(1)$ constant that is supposed to be determined
by the UV physics. We assume this relation for gaugino masses 
throughout this paper.\footnote{We however note that our discussion
is less affected even though we take a different value of 
$M_1/M_3$, since the contribution of the bino mass to RGEs is
relatively small as the U(1)$_Y$ gauge coupling is smaller 
than the other gauge couplings.}
It is readily found from Eq.~\eqref{eq:gauginomasses} that $M_1$ is
always larger than $M_2$ for $M_2/M_3 > 1$, and in particular $M_1
\simeq 2 M_2$ for $M_2/M_3 \simeq 3$. 
From Fig.~\ref{fig:r23vsms}, we find that a small
change in $M_2/M_3$ drastically affects the GUT-scale value of
$m_S^2$. For a given value of $M_2/M_3$, the universality condition
$m_S^2 = m_0^2$ can be satisfied by taking an appropriate value of
$\tan\beta $---there may be two different choices of $\tan\beta$ that
give $m_S^2 = m_0^2$. The required values of $\tan\beta$ tend to get
larger for a larger $M_2/M_3$. We however note that the condition $m_S^2
= m_0^2$ is not the only one needed to be satisfied; other
conditions such as the stability conditions discussed above should also
be satisfied, and in fact these requirements disfavor a small value of
$M_2/M_3$ as we see below.

A non-universal gaugino mass spectrum is advantageous also for the
vacuum stability. Since a smaller $|m_{H_u}^2|$, which can be realized
with a large wino mass, results in a larger value of
$V_{S,D}^{(\text{min})}$ in Eq.~\eqref{eq:vsdmin}, the desired vacuum
can easily be deeper than this unwanted minimum. In addition, a large
wino mass tends to make the right-hand side in
Eq.~\eqref{eq:dfflatlocalcond} large compared to the left-hand side,
which allows the $\phi$ direction to be stabilized. 
In our parameter scanning, \texttt{NMSSMTools 5.3.1} only checks 
whether there is a deeper minimum along the directions 
that one of the three fields, $H_u$, $H_d$ and $S$, has a non-zero 
field value. We expect that deeper minima along the $F$-flat 
and/or the $D$-flat directions are absent 
in the parameter space where the $M_2/M_3$ is large enough to 
achieve the correct EWSB vacuum.

Another obstacle to a viable parameter point in the CNMSSM is stau being the
LSP or tachyonic, for which the non-universal gaugino mass conditions $M_1/M_3,
M_2/M_3 > 1$ are again helpful. In particular, a larger value of $M_1$
increases the soft mass of the right-handed stau at low energies through
the RG effect. This prevents stau from being the LSP or tachyonic even
for a large $|A_0|$.

\begin{figure}
\centering
\subcaptionbox{\label{fig:err0}
$\text{sign}(v_s) = -$, $A_0 = 0$
}
{\includegraphics[width=0.49\textwidth]{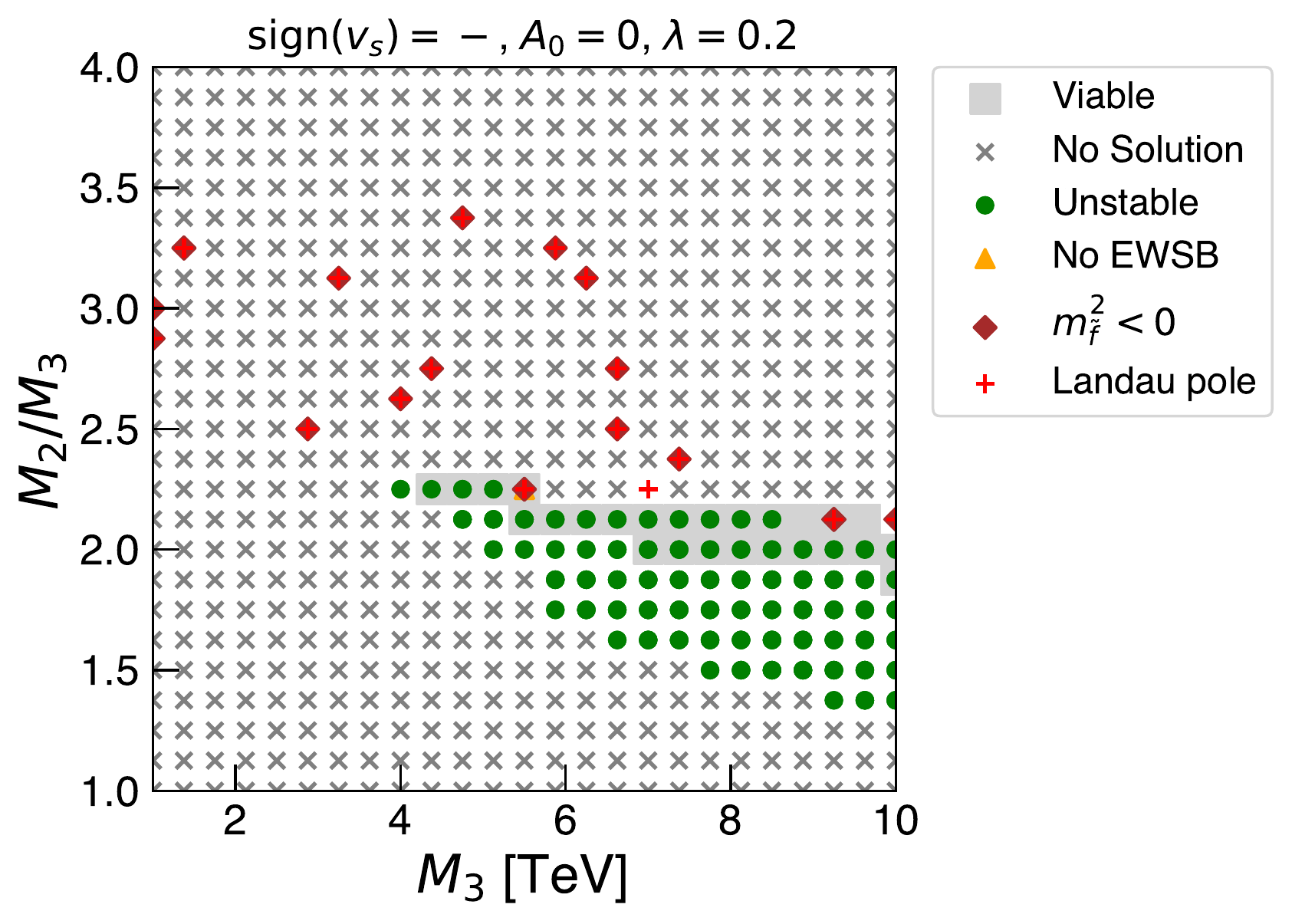}}
\subcaptionbox{\label{fig:err0vspl}
$\text{sign}(v_s) = +$, $A_0 = 0$
}
{\includegraphics[width=0.49\textwidth]{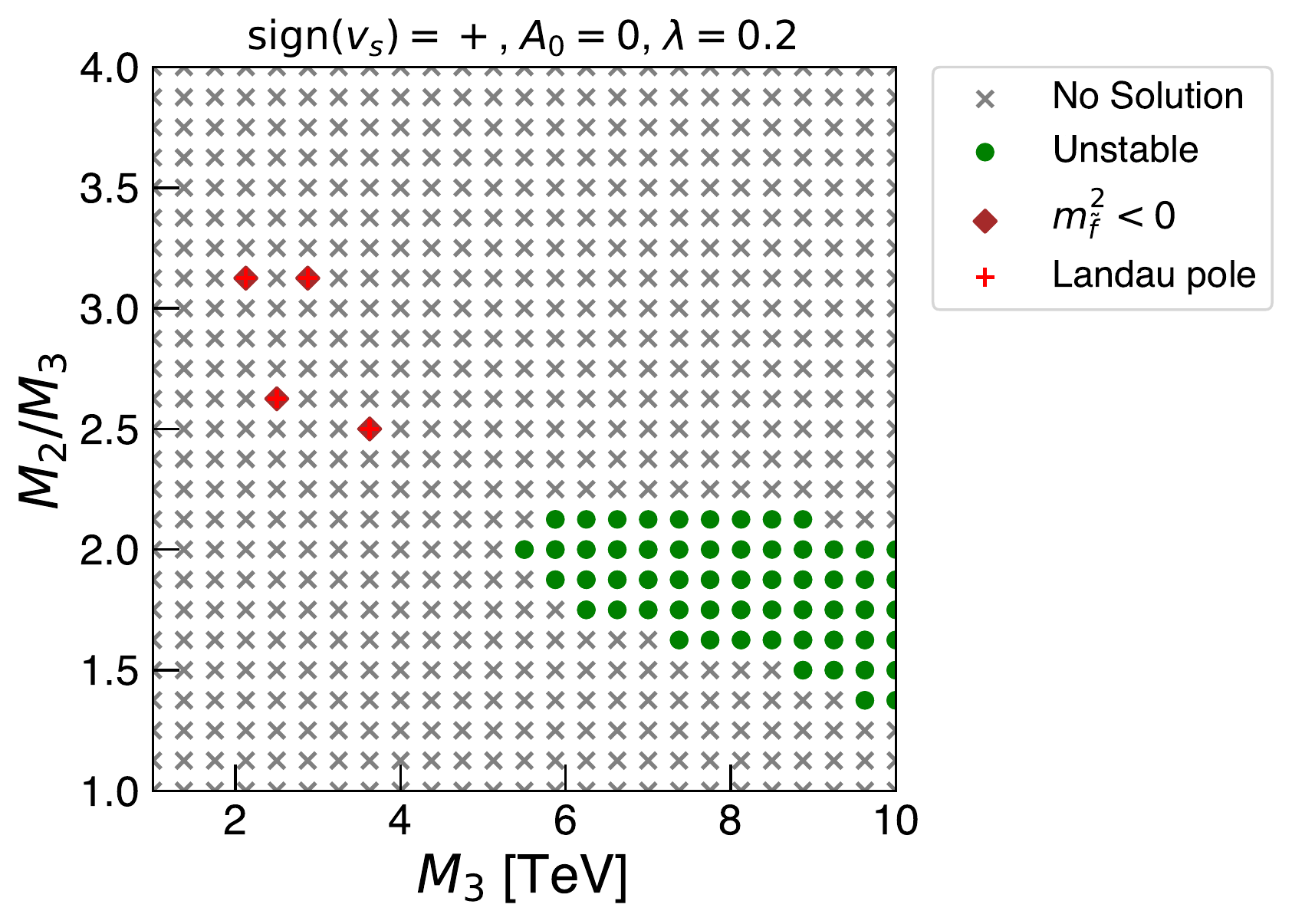}} \\[3pt]
\subcaptionbox{\label{fig:errpl}
$\text{sign}(v_s) = -$, $A_0 = M_3$
}
{\includegraphics[width=0.49\textwidth]{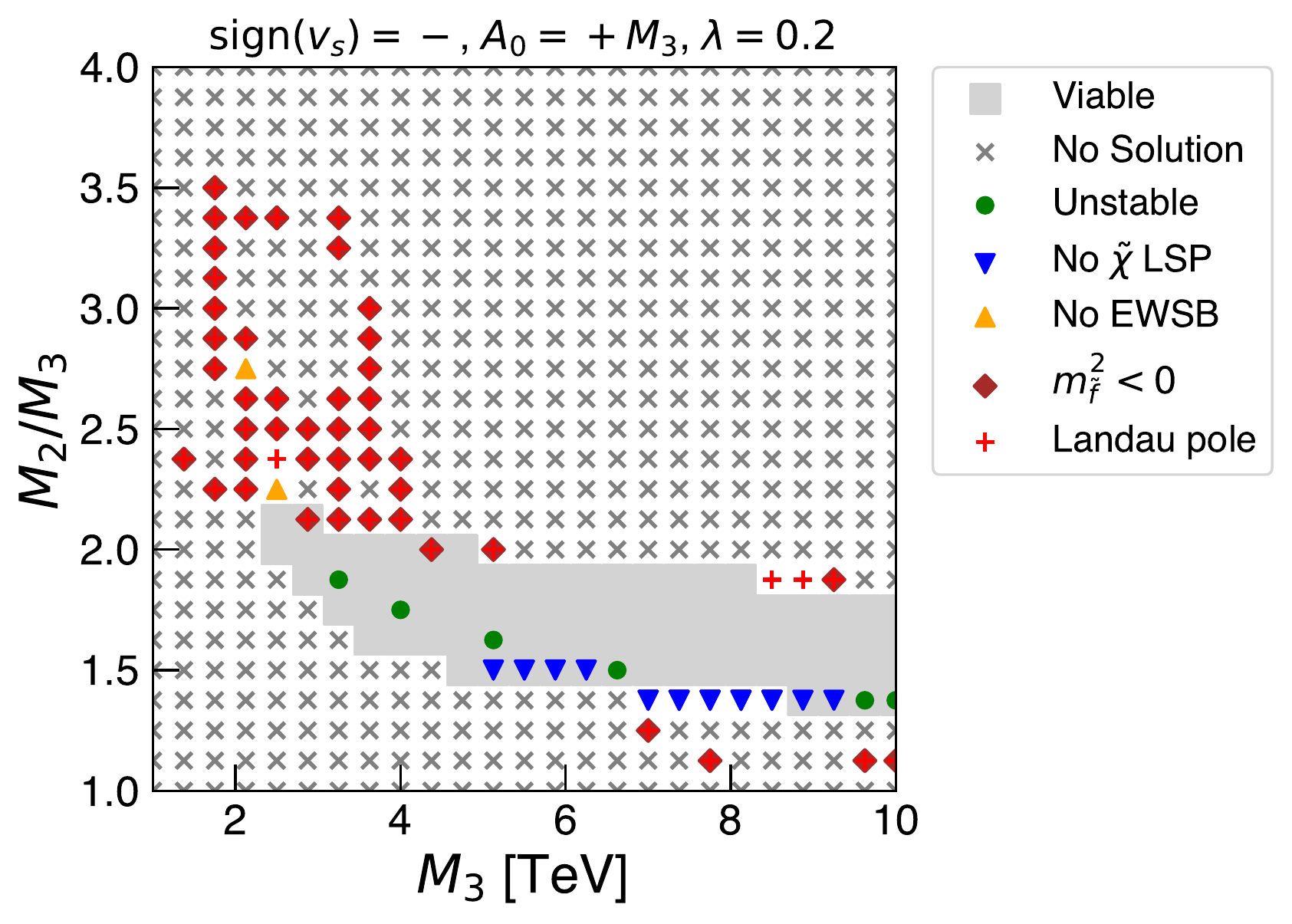}}
\subcaptionbox{\label{fig:errplvspl}
$\text{sign}(v_s) = +$, $A_0 = M_3$
}
{\includegraphics[width=0.49\textwidth]{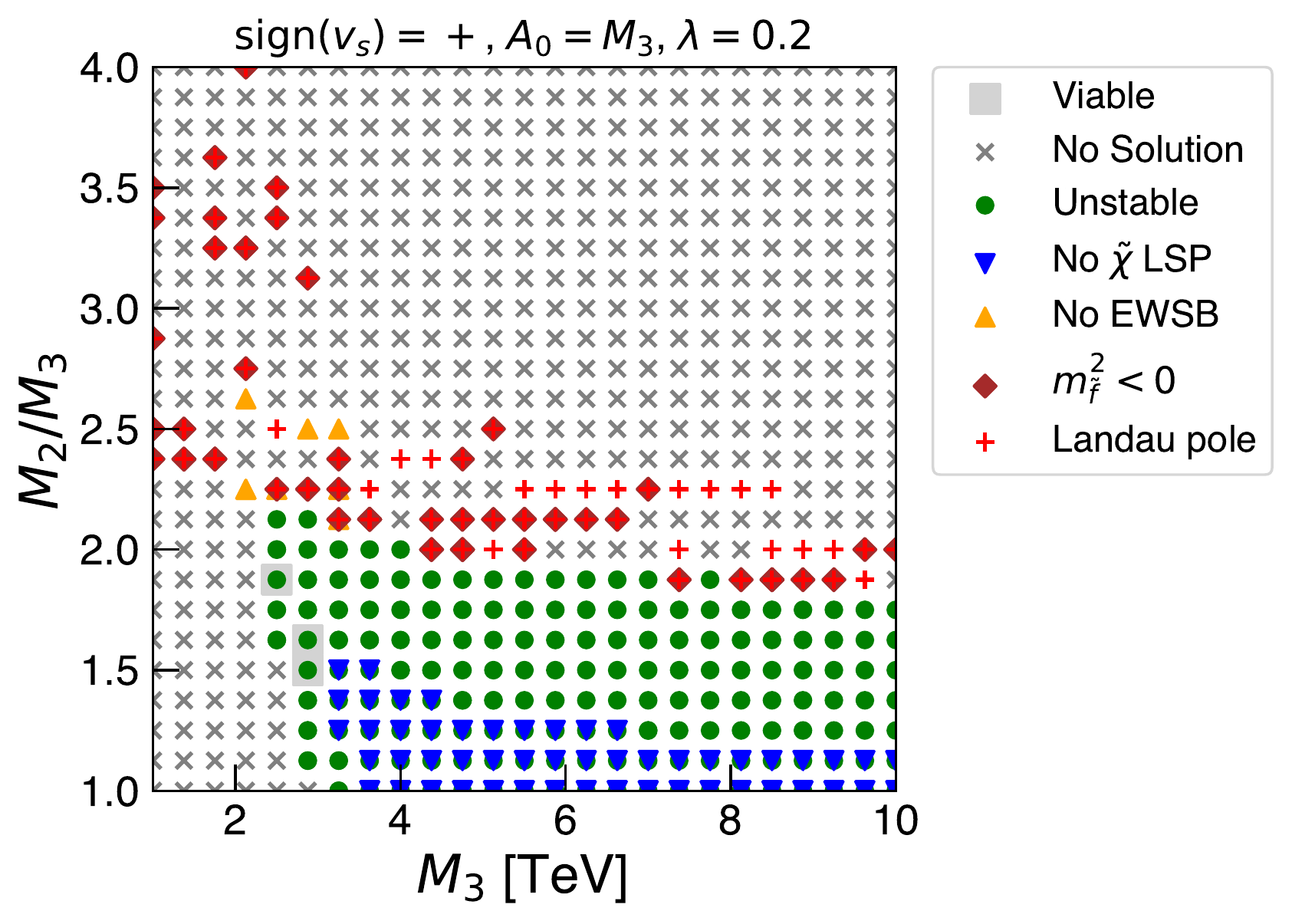}}  \\[3pt]
\subcaptionbox{\label{fig:errmi}
$\text{sign}(v_s) = -$, $A_0 = -M_3$
}
{\includegraphics[width=0.49\textwidth]{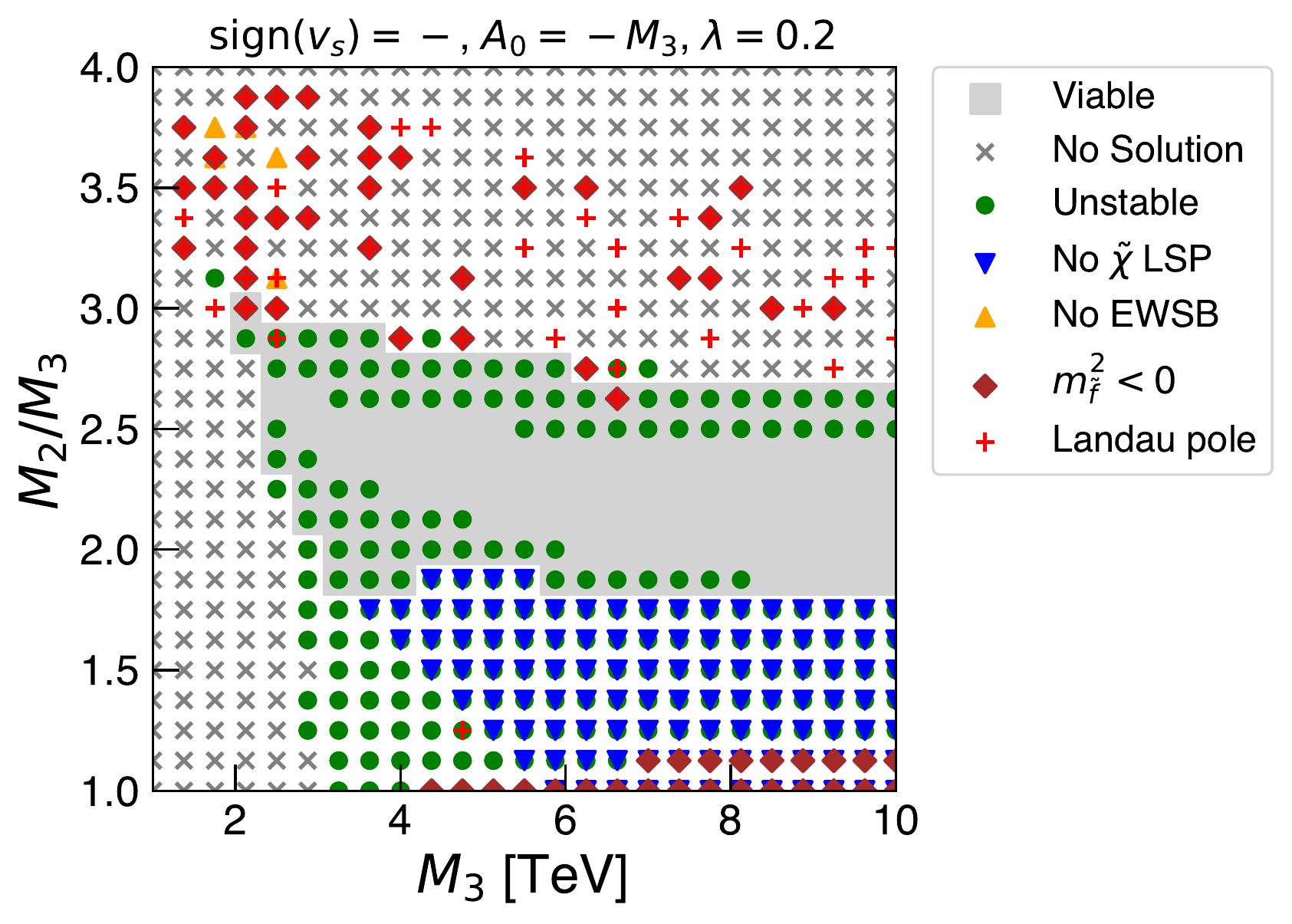}}
\subcaptionbox{\label{fig:errmivspl}
$\text{sign}(v_s) = +$, $A_0 = -M_3$
}
{\includegraphics[width=0.49\textwidth]{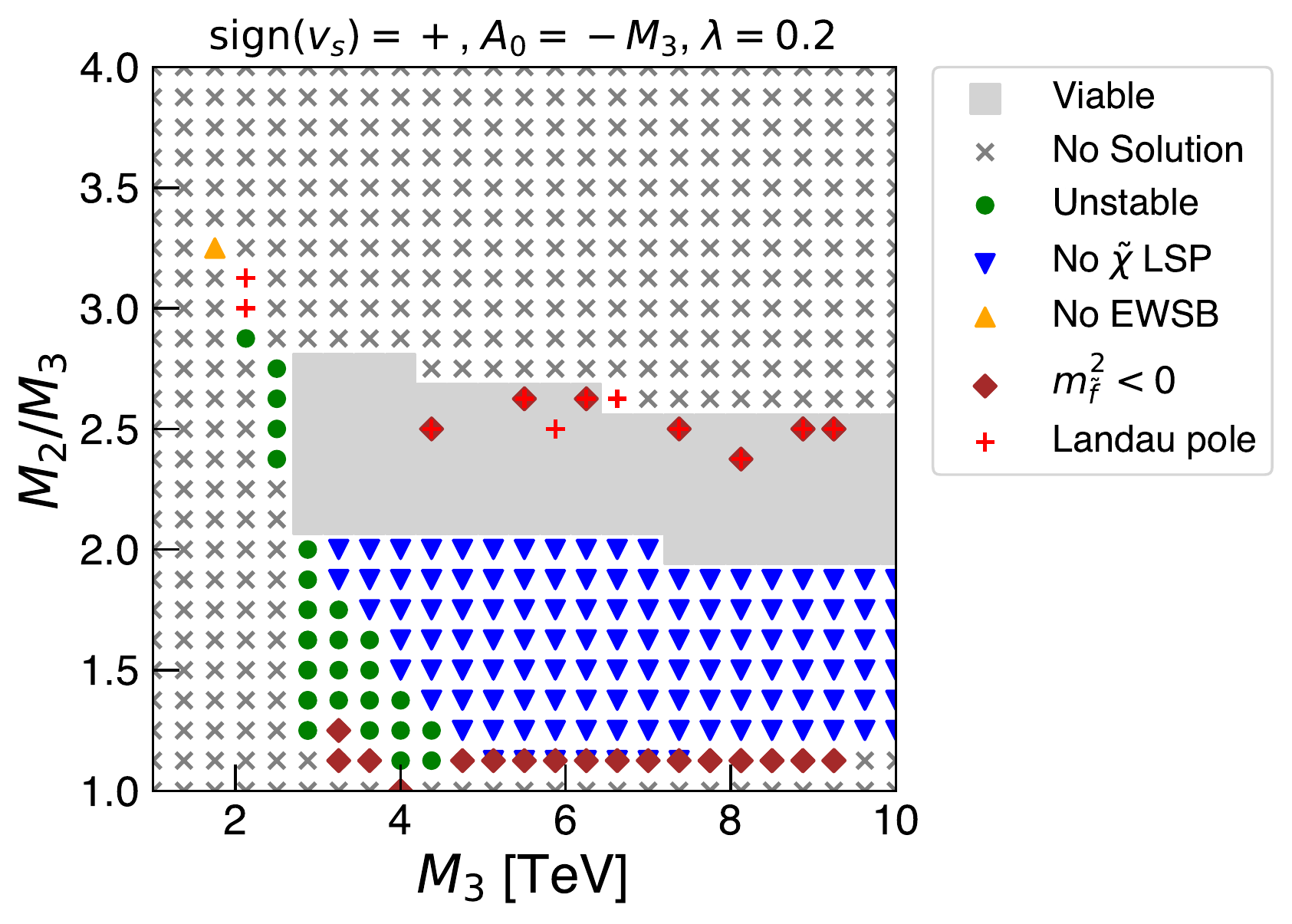}}
\caption{Viable/excluded parameter points in the $M_3$-$M_2/M_3$ plane
 for each choice of $\text{sign} (v_s)$ and $A_0$. Here we set $m_0 =
 1$~TeV and $\lambda = 0.2$, and determine $\tan\beta$ from the
 universality condition $m_S^2 = m_0^2$.
}
\label{fig:err}
\end{figure}

In Fig.~\ref{fig:err}, we show viable and excluded parameter points in
the $M_3$-$M_2/M_3$ plane for each choice of $\text{sign} (v_s)$ and
$A_0$. Here we set $m_0 = 1$~TeV and $\lambda = 0.2$, and determine
$\tan\beta$ from the universality condition $m_S^2 = m_0^2$. The viable
parameter region is shown in the gray shaded area, while other marks
indicate that the corresponding parameter points are excluded for
various reasons. There are
points at which several solutions for $\tan\beta$ exist, for which the
marks associated to these solutions are
overlapped. In Fig.~\ref{fig:err0}, where we set $\text{sign} (v_s)$
to be negative and $A_0 = 0$, there is no solution
compatible with $m_S^2 = m_0^2$ for $M_1 = M_2 = M_3$,\footnote{We see
from Eq.~\eqref{eq:gauginomasses} that if $M_2 = M_3$ then $M_1 = M_3$,
\textit{i.e.}, the universal gaugino mass is obtained. }
as indicated by the gray crosses. This is consistent with our discussion
in Sec.~\ref{sec:CMSSM}. Nevertheless, if we allow non-universal gaugino masses, 
we can then find viable parameter points, though they
are restricted to a narrow strip spreading at $M_3 \gtrsim 4$~TeV and $M_2/M_3
\simeq 2$. If we flip the sign of $v_s$, on the other hand, there is no
viable parameter point as shown in Fig.~\ref{fig:err0vspl}. The reason
is as follows. If we take $A_0 = 0$ and $M_2> M_3$, the low-scale value
of $A_\kappa$ ($A_\lambda$) tends to be positive (negative), as implied
by Eq.~\eqref{eq-AkB} (Eq.~\eqref{eq-AlB}). Given $v_s > 0$, {\it i.e.},
$\mu_{\text{eff}} > 0$, $B_{\text{eff}} = A_\lambda + \kappa
v_s/\sqrt{2} > 0$ is required in order for the vacuum condition
\eqref{eq:vaccond2} to be satisfied. This is possible only if $\kappa
v_s >0$. However, if $A_\kappa > 0$ and $\kappa v_s > 0$, then the
pseudo-scalar mass squared becomes negative as seen from
Eq.~\eqref{eq:ma2}. This means that the vacuum we consider is actually
unstable, which is indicated by the green dots in
Fig.~\ref{fig:err0vspl}. The narrow strip observed in
Fig.~\ref{fig:err0} is extended if we take a large $|A_0|$, as seen in
Figs.~\ref{fig:errpl} and \ref{fig:errmi}, where we set $A_0 = +M_3$ and
$-M_3$, respectively. Now the viable parameter region is considerably
extended, since a large $|A_0|$ makes it easy to obtain a large value of
$v_s$ given by Eq.~\eqref{eq:vssol} and thus to satisfy the condition
\eqref{eq:simpvaccond1}. As a side effect, a large $|A_0|$ may result in a
light (or tachyonic) stau as discussed above; we find such points
indicated by blue triangles, where the lightest neutralino is not the
LSP, for a relatively small $M_2/M_3$. As we see, the
light/tachyonic-stau problem is evaded for a larger value of $M_2/M_3$.
For a positive $v_s$, viable parameter points are still rarely
found for $A_0 = + M_3$ as shown in Fig.~\ref{fig:errplvspl} due to 
similar reasoning given above. For $A_0 = -M_3$, on the other hand, we
find viable parameter region for $M_2/M_3 \sim 2.5$, as seen in
Fig.~\ref{fig:errplvspl}. Motivated by this observation---namely,
$\text{sign}(v_s) = -$ allows larger number of viable parameter points,
we always take $\text{sign}(v_s) =-$ in the following analysis.

All in all, the restrictions from the tadpole conditions, the vacuum
stability conditions, and the light/tachyonic stau can significantly be
alleviated with non-universal gaugino masses. This opens up
new viable parameter regions where phenomenological consequences are
quite different from those in the CNMSSM, as we see below.

\section{Phenomenological implications} 
\label{sec:pheno}

\begin{table}[t!]
 \begin{center}
\caption{Typical mass spectrum of the NMSSM with non-universal gaugino
  masses. }
\label{tab:ex1}
\vspace{5pt}
\begin{tabular}{ll|ll|ll}
\hline
\hline
\rowcolor{LightGray}
\multicolumn{6}{c}{GUT-scale input parameters ($M_{\text{GUT}} =
 7.57\times 10^{15}$~GeV)} \\
\hline
$m_0$ & 1~TeV & $A_0$ & $-5.792$~TeV &  &  \\
$M_1$ & $28.67$~TeV & $M_2$ & $15.32$~TeV & $M_3$ & $5.792$~TeV \\
\hline
\rowcolor{LightGray}
\multicolumn{6}{c}{SUSY-scale input parameters} \\
\hline
$\text{sign} (v_s) $ & $-$ & $\lambda$ & $0.207$  &  &  \\
\hline
\rowcolor{LightGray}
\multicolumn{6}{c}{Output parameters} \\
\hline
$\tan\beta$ & $54.61$ &$\kappa$ & $-0.455$ & $\mu_{\text{eff}}$ &
		     $-1.049$~TeV \\
$A_\lambda$ & $-2.548$~TeV & $A_\kappa$ & $-2.668$~TeV &$m_S^2$ & $-
		     4.475 $~TeV$^2$
\\
$M_1$ & $13.73$~TeV & $M_2$ & $12.75$~TeV & $M_3$ & $10.88$~TeV \\
\hline
\rowcolor{LightGray}
\multicolumn{6}{c}{Mass spectrum} \\
\hline
$m_{h_1}$ & $124.3$~GeV & $m_{h_2}$& $3.092$~TeV &$m_{h_3}$ &
		     $3.854$~TeV \\
$m_{a_1}$ & $3.091$~TeV & $m_{a_2}$& $4.224$~TeV &$m_{H^\pm}$ &
		     $3.093$~TeV \\
$m_{\widetilde{\chi}^0_1}$ & $1.077$~TeV & $m_{\widetilde{\chi}^\pm_1}$&
	     $1.078$~TeV &$m_{\widetilde{g}}$ & $11.47$~TeV \\
$m_{\widetilde{t}_1}$ & $7.685$~TeV & $m_{\widetilde{b}_1}$& $6.776$~TeV
	     &$m_{\widetilde{\tau}_1}$ & $7.620$~TeV \\
\hline
\multicolumn{2}{l}{Other sfermions:} & 
\multicolumn{2}{l|}{$9.4$--$13.1$~TeV} && \\
\hline
\rowcolor{LightGray}
\multicolumn{6}{c}{Dark matter} \\
\hline
$\Omega_{\text{LSP}} h^2$ & $0.112$ & $\sigma_{\text{SI}}^{(p)}$ & $1.8
	     \times 10^{-47}~\text{cm}^2$  & $\sigma_{\text{SD}}^{(p)}$ & $1.1
	     \times 10^{-45}~\text{cm}^2$ \\
\hline
\rowcolor{LightGray}
\multicolumn{6}{c}{Couplings at the GUT scale} \\
\hline
$g_1$ & $0.685$ & $g_2$ & $0.685$ & $g_3$ & $0.688$ \\
$y_t$ & $0.574$ & $y_b$ & $0.498$ & $y_\tau$ & $0.651$ \\
$\lambda$ & $0.302$ & $\kappa$ & $-0.656$ & & \\
\hline
\hline
\end{tabular}
 \end{center}
\end{table}

Now we discuss the phenomenology of the NMSSM with non-universal gaugino
masses. First, we show in Table~\ref{tab:ex1}
a typical mass spectrum of this scenario which is consistent with $m_h \simeq 125$~GeV
and $\Omega_{\text{DM}} h^2 \simeq 0.12$. 
Here, all of the output parameters except for $\tan \beta$ are evaluated
at the SUSY scale, while $\tan\beta$ is given at the $Z$-boson mass
scale. This mass spectrum is computed again with \texttt{NMSSMTools
5.3.1} \cite{Ellwanger:2004xm, Ellwanger:2005dv, Ellwanger:2006rn},
while the thermal relic abundance of the LSP, $\Omega_{\text{LSP}} h^2$,
and its spin-independent and spin-dependent scattering cross sections
with proton, $\sigma_{\text{SI}}^{(p)}$ and $\sigma_{\text{SD}}^{(p)}$,
respectively, are obtained by using \texttt{MicrOMEGAs}
\cite{Belanger:2005kh} implemented in \texttt{NMSSMTools 5.3.1}. For the
computation of $\sigma_{\text{SI}}^{(p)}$, we use the latest compilation
for the nucleon scalar matrix elments given in
Ref.~\cite{Ellis:2018dmb}: $f_{T_u}^{(p)} = 0.018 (5)$, $f_{T_d}^{(p)} =
0.027 (7)$, $f_{T_s}^{(p)} = 0.037 (17)$, $f_{T_c}^{(p)} = 0.078 (2)$,
$f_{T_b}^{(p)} = 0.072 (2)$, and $f_{T_t}^{(p)} = 0.069 (1)$, where
the heavy quark matrix elements are obtained from those for light quarks
through an ${\cal O}(\alpha_s^3)$ perturbative QCD calculation. We have
checked that the use of these matrix elements results in an enhancement
in $\sigma_{\text{SI}}^{(p)}$ by about 13\% compared with that computed
with the default values of the nucleon matrix elements exploited in
\texttt{MicrOMEGAs}. For $\sigma_{\text{SD}}^{(p)}$, we use the default
values of the matrix elements in \texttt{MicrOMEGAs}. We also note that
the computation of the Higgs masses in the NMSSM suffers from a large
theoretical uncertainty; indeed, various public codes predict values of
$m_h$ differing by as large as a few GeV \cite{Staub:2015aea}. 
We use the option \texttt{2} in \texttt{NMSSMTools
5.3.1} to compute the Higgs masses, which are found to be in good
agreement with the results of other codes as shown in
Ref.~\cite{Staub:2015aea}. 

In Table~\ref{tab:ex1}, we take $M_2 \simeq 2.5 \times M_3$ and $M_1
\simeq 2\times M_2$ to realize a non-universal gaugino mass spectrum consistent
with the relation \eqref{eq:gauginomasses}. 
As seen in the table, $\kappa$ is negative and its absolute
value is larger than $\lambda$, which is again a generic feature as we
will see below. Our choice of a negative $A_0$ leads to a negative value
of $A_\kappa$ at low energies, which requires a negative $\kappa$ to
make $m_a^2$ in Eq.~\eqref{eq:ma2} positive since $v_s$ is taken to be
negative. As $\lambda < |\kappa|$, higgsino is lighter than singlino,
and in fact the neutral higgsino is the LSP in this mass spectrum. Its
mass is $\simeq 1$~TeV, with which its thermal relic abundance agrees to
the observed DM density. In spite of the universality condition $m_S^2 =
m_0^2 > 0$ at the GUT scale, $m_S^2$ at the SUSY scale is
negative---thus, the generation of a non-zero VEV of the singlet field
is induced radiatively. This is a distinct feature of
this setup compared with the CNMSSM, where $m_S^2$
scarcely runs and thus is always positive. The gaugino masses at the
SUSY scale are close to each other, similarly to those in the mirage
mediation \cite{Choi:2004sx, Choi:2005ge, Endo:2005uy, Choi:2005uz,
Choi:2005hd, Kitano:2005wc, Choi:2006xb}. The lightest neutral Higgs
boson corresponds to the SM-like Higgs boson, while the other Higgs
bosons have masses of a few TeV. The third-generation sfermions are
relatively light compared with the other sfermions, but they are much
heavier than the LSP. All of the couplings are found to remain
perturbative up to the GUT scale $M_{\text{GUT}} \simeq 8 \times
10^{15}$~GeV, which is relatively low compared with the typical
GUT scale ($\simeq 2\times 10^{16}$~GeV) in many SUSY mass
spectra. This is basically due to large values of gaugino masses
\cite{Hisano:2013cqa}, and may result in a large rate of the 
dimension-six proton decay induced by the exchange of GUT gauge bosons.

Next, we perform a parameter scan to search for mass spectra that
are consistent with the observed value of the Higgs mass as well as the
DM density. We fix/scan the input parameters as follows:
\begin{itemize}
 \item $m_0 = 1$~TeV.
 \item $1~\text{TeV} < M_3 < 12.5$~TeV. 
 \item $1.5 < M_2/M_3 < 4$. 
 \item $0 < |A_0| < 3 M_3$.
 \item $0.01 < \lambda < 0.5$.
 \item $\text{sign}(v_s) = -$.
\end{itemize}
We then check that each parameter point satisfies the required
conditions discussed in Sec.~\ref{sec:conds}.

\begin{figure}[t]
\centering
\subcaptionbox{\label{fig:lamkapmi}
$A_0 < 0$
}
{\includegraphics[width=0.49\textwidth]{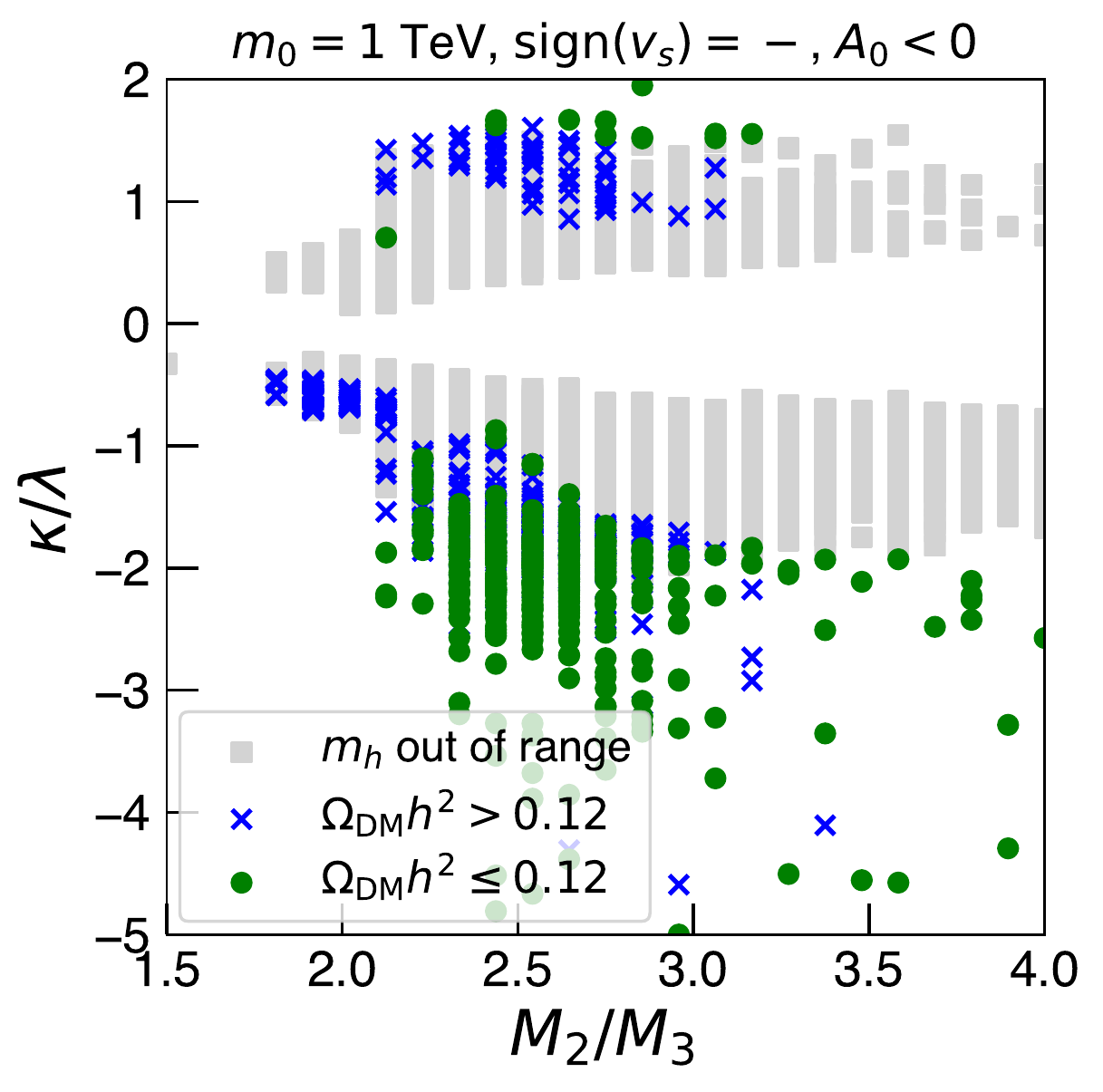}}
\subcaptionbox{\label{fig:lamkappl}
$A_0 > 0$
}
{\includegraphics[width=0.49\textwidth]{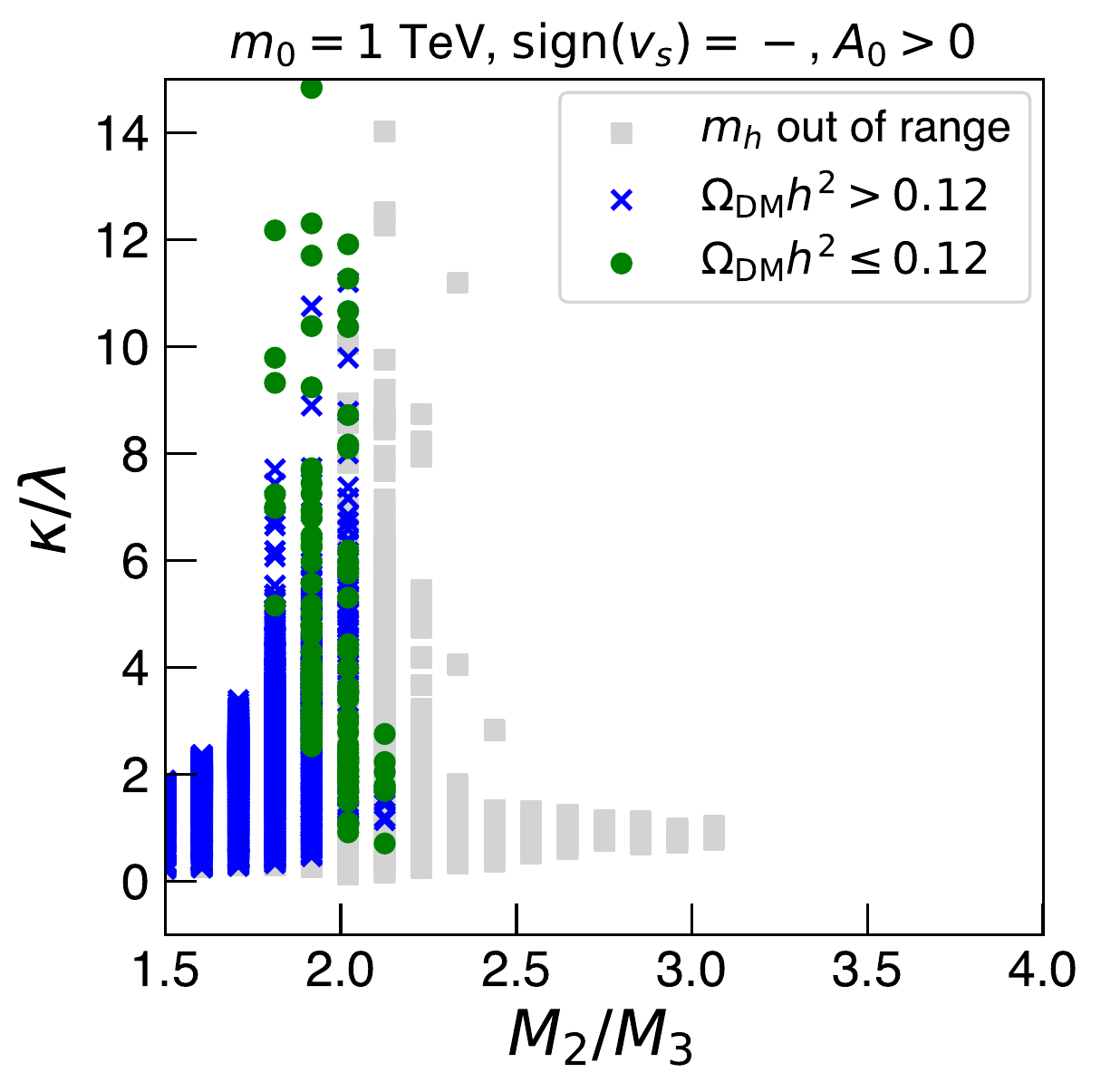}}
\caption{Scatter plots of $\kappa/\lambda$ against $M_2/M_3$. The green
 dots and blue crosses represent the points where the relic abundance of
 the neutralino LSP is below and above the observed value,
 respectively. The points indicated by gray squares predict $m_h$ that
 is out of the acceptable mass range: $122.1~\text{GeV} < m_h <
 128.1$~GeV. Here we fix $m_0 = 1$~TeV and $\text{sign}(v_s) = -$.
}
\label{fig:lamkap}
\end{figure}

In Fig.~\ref{fig:lamkap}, we show scatter plots of $\kappa/\lambda$
against $M_2/M_3$, where Fig.~\ref{fig:lamkapmi} and
Fig.~\ref{fig:lamkappl} are for $A_0 < 0$ and $A_0 > 0$, respectively. 
The green dots (blue crosses) represent the points
where the relic abundance of the neutralino LSP is below (above) the
observed value. These parameter points are also required
to reproduce the mass of the SM-like Higgs boson $m_h
\simeq 125$~GeV; with the theoretical uncertainty in the
computation of $m_h$ taken into account, we regard
$122.1~\text{GeV} < m_h < 128.1$~GeV as the allowed range for 
$m_h$, following the prescription in \texttt{NMSSMTools}. The gray
squares in Fig.~\ref{fig:lamkap} correspond to the points where 
$m_h$ is out of this range. From these plots, we find
that the viable parameter points spread over $2 \lesssim M_2/M_3
\lesssim 4$ in the case of a negative $A_0$, while they are localized at
$M_2/M_3 \sim 2$ for a positive $A_0$. 
$|\kappa|$ tends to be larger than $\lambda$ for $A_0 < 0$,
especially for those fall into the allowed range of the Higgs boson
mass. For $A_0 > 0$, we find some points that predict $|\kappa| <
\lambda$, but such points are disfavored due to the overproduction of the
LSP. These results imply that higgsino is lighter than singlino in most
of the viable parameter points and the LSP is the higgsino-like
neutralino. In addition, we find more points that give a negative
$\kappa$ than those yield a positive $\kappa$ for $A_0<0$, while for
$A_0>0$ most of the points predict a positive $\kappa$. This is because
the sign of $\kappa$ should be equal to the sign of $A_\kappa$ at the
SUSY-breaking scale, given $v_s < 0$ and the condition \eqref{eq:ma2}. We also note
that there are quite a few points with $\kappa/\lambda < -5$ ($\kappa/\lambda > 15$) for
$A_0 < 0$ ($A_0 > 0$), which are out of the region shown in this figure.

\begin{figure}[t]
\centering
\subcaptionbox{\label{fig:mgvsr23mi}
$A_0 < 0$
}
{\includegraphics[width=0.49\textwidth]{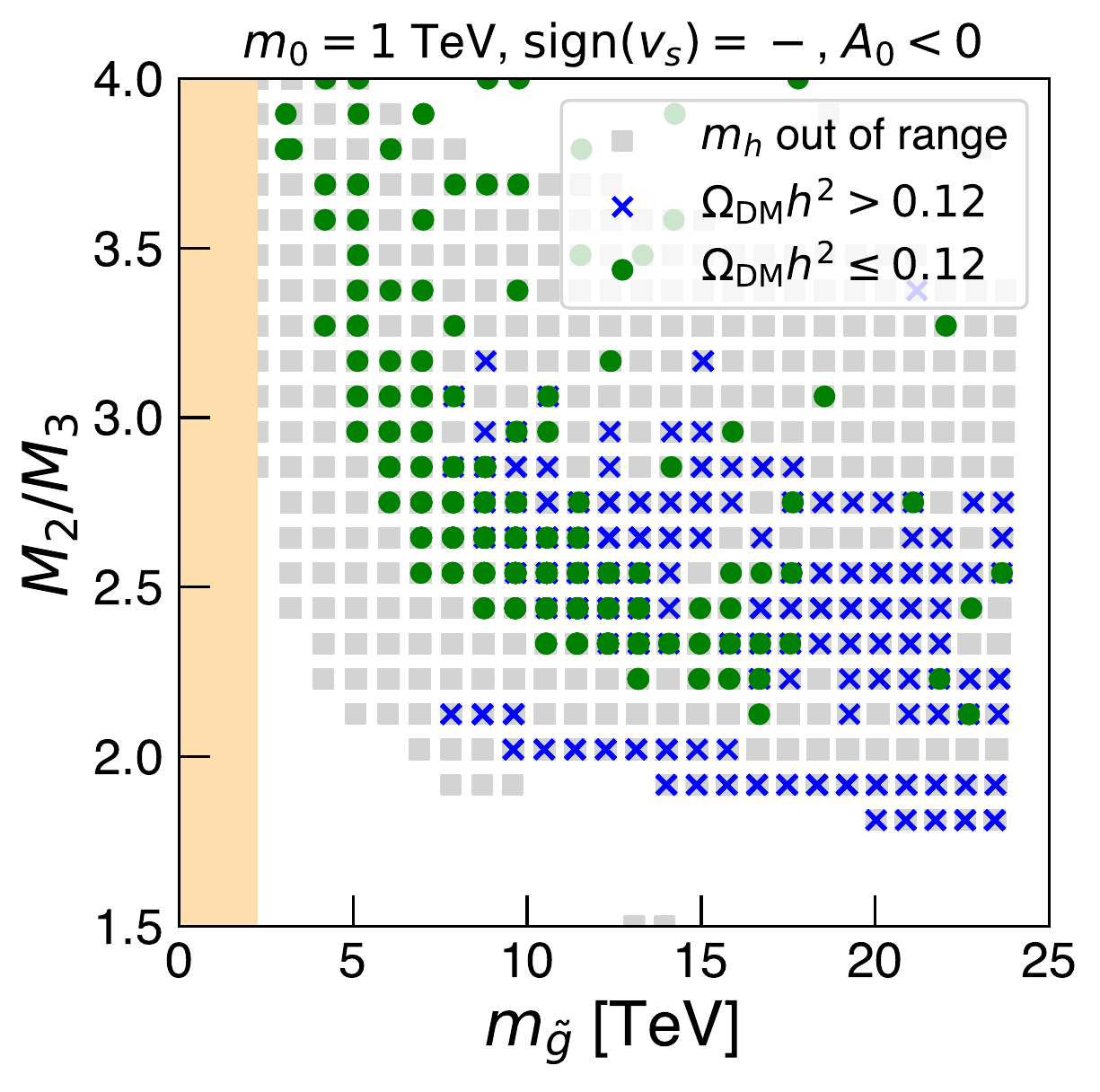}}
\subcaptionbox{\label{fig:mgvsr23pl}
$A_0 > 0$
}
{\includegraphics[width=0.49\textwidth]{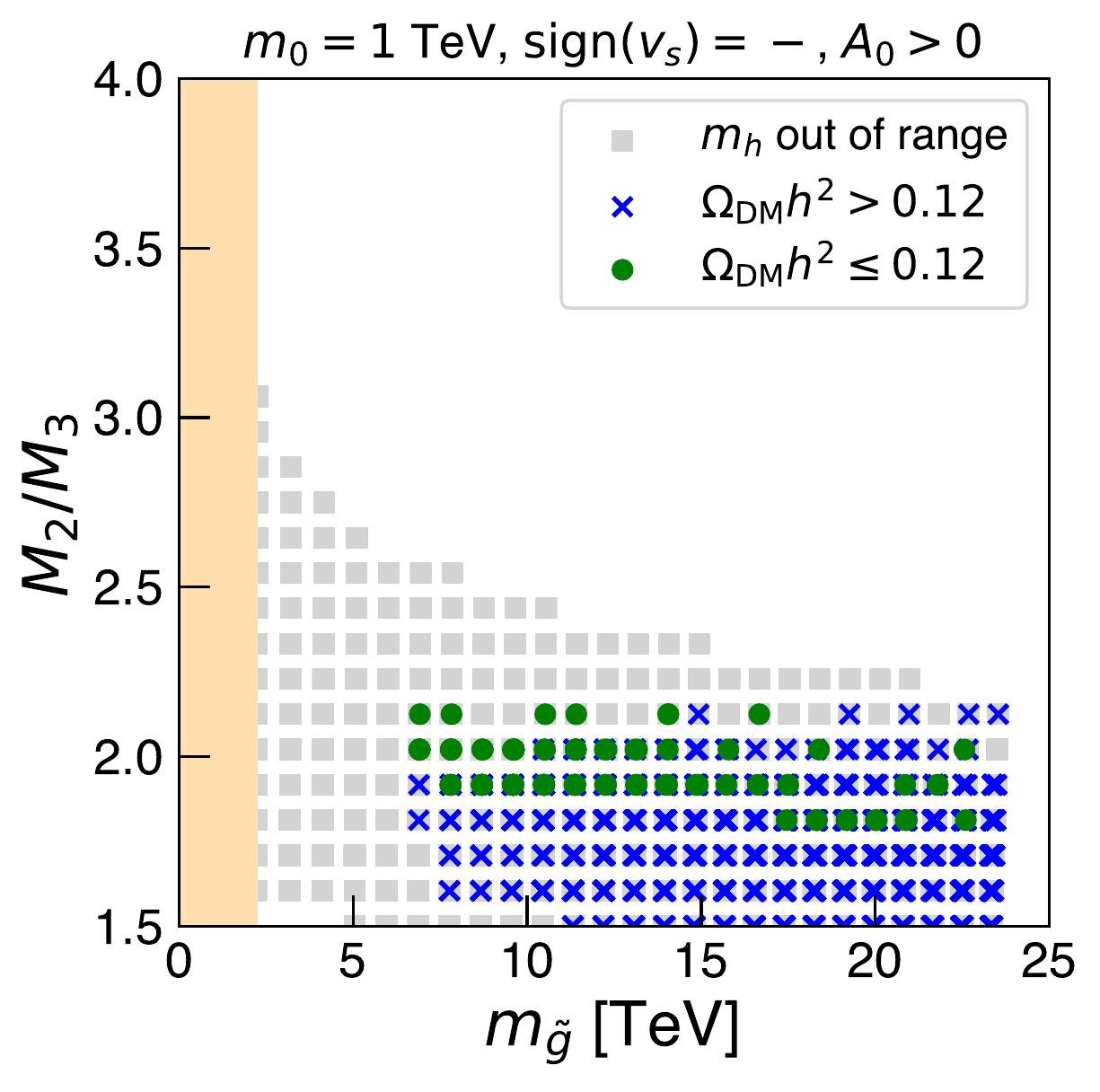}}
\caption{Scatter plots in the $m_{\widetilde{g}}$ -$M_2/M_3$ plane.
 The meaning of the marks and the choice of $m_0$ and $\text{sign}(v_s)$
 are the same as in Fig.~\ref{fig:lamkap}. The orange shaded region
 represents the current ATLAS bound on the gluino mass,
 $m_{\widetilde{g}} \gtrsim 2.2$~TeV \cite{ATLAS:2018yhd} for the LSP
 lighter than $\sim 1$~TeV. 
}
\label{fig:mgvsr23}
\end{figure}

Next we show the same parameter points as above in the
$m_{\widetilde{g}}$ -$M_2/M_3$ plane in Fig.~\ref{fig:mgvsr23}, where the
meaning of the marks is the same as in Fig.~\ref{fig:lamkap}. The orange
shaded region represents the latest ATLAS bound on the gluino mass,
$m_{\widetilde{g}} \gtrsim 2.2$~TeV \cite{ATLAS:2018yhd} for the LSP lighter than $\sim 1$~TeV.\footnote{
For the results of the ATLAS and the CMS with the fewer data; 
see, \textit{e.g.}, Refs.~\cite{Aaboud:2017vwy,Aaboud:2017hrg, Sirunyan:2018vjp}. 
} 
These plots show that most of the parameter points consistent with the DM
abundance predict the gluino mass to be $3~\text{TeV} \lesssim
m_{\widetilde{g}}  \lesssim 20$~TeV ($7~\text{TeV} \lesssim
m_{\widetilde{g}}  \lesssim 23$~TeV) for $A_0 < 0$ ($A_0 > 0$). As a
comparison, we note that a future 100~TeV $pp$ collider may probe
gluinos with a mass of $\lesssim 13$~TeV \cite{Cohen:2013xda,
Ellis:2015xba, Arkani-Hamed:2015vfh, Golling:2016gvc}; thus, many
parameter points in the non-universal gaugino mass scenario may be
probed in the future, especially if $M_2/M_3 \gtrsim 3$. It is also
found that other colored particles such as stop and sbottom may be
within the reach of a 100~TeV collider as well for a part of the viable
parameter points.

As we see in the above plots, many parameter points give the LSP whose
thermal relic abundance is within the observed DM density. As seen from
Fig.~\ref{fig:lamkap}, in this case, $\lambda < |\kappa|$ and thus
higgsino is lighter than singlino. Since gaugino masses are rather
heavy in our scenario, the LSP in the viable parameter points is always
higgsino-like neutralino, and its thermal relic abundance,
$\Omega_{\text{LSP}} h^2$, is smaller than the observed value
$\Omega_{\text{DM}} h^2 \simeq 0.12$ if its mass is $\lesssim
1$~TeV. Higgsino has couplings with the Higgs bosons via the mixing
with singlino, bino and wino, and through these couplings it can scatter
off nucleons. According to Fig.~\ref{fig:lamkap}, 
$|\kappa|/\lambda$ is ${\cal O}(1)$ for most of the viable parameter points, and thus
the singlino mass is expected to be of the same order as the higgsino
mass. In this case, the higgsino-singlino mixing is sizable, and we
expect a detectable value of the LSP-nucleon scattering cross
section. If this is the case, we may probe the LSP in the future DM
direct detection experiments.

\begin{figure}[t]
\centering
\subcaptionbox{\label{fig:sigsimi}
$A_0 < 0$
}
{\includegraphics[width=0.49\textwidth]{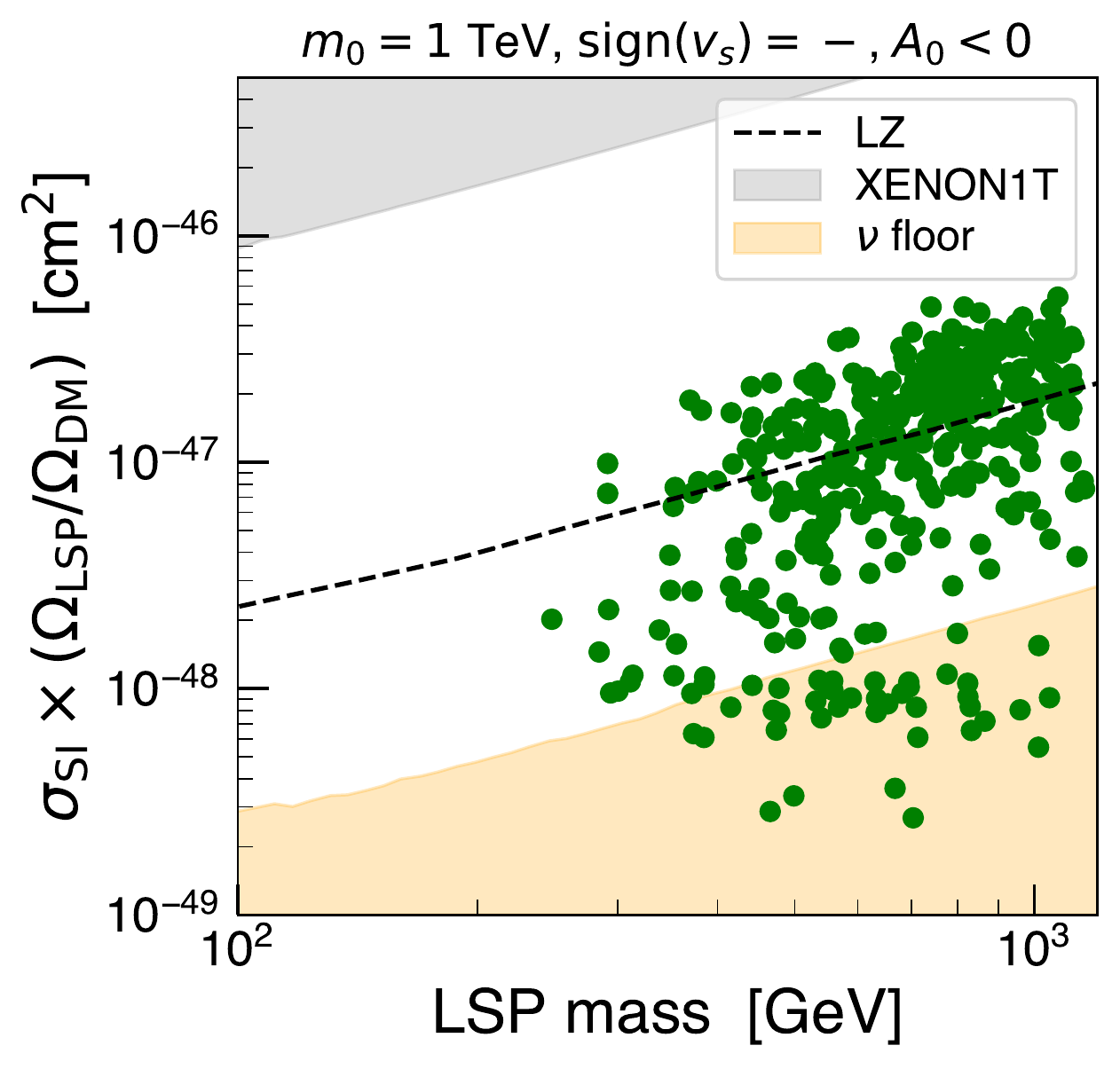}}
\subcaptionbox{\label{fig:sigsipl}
$A_0 > 0$
}
{\includegraphics[width=0.49\textwidth]{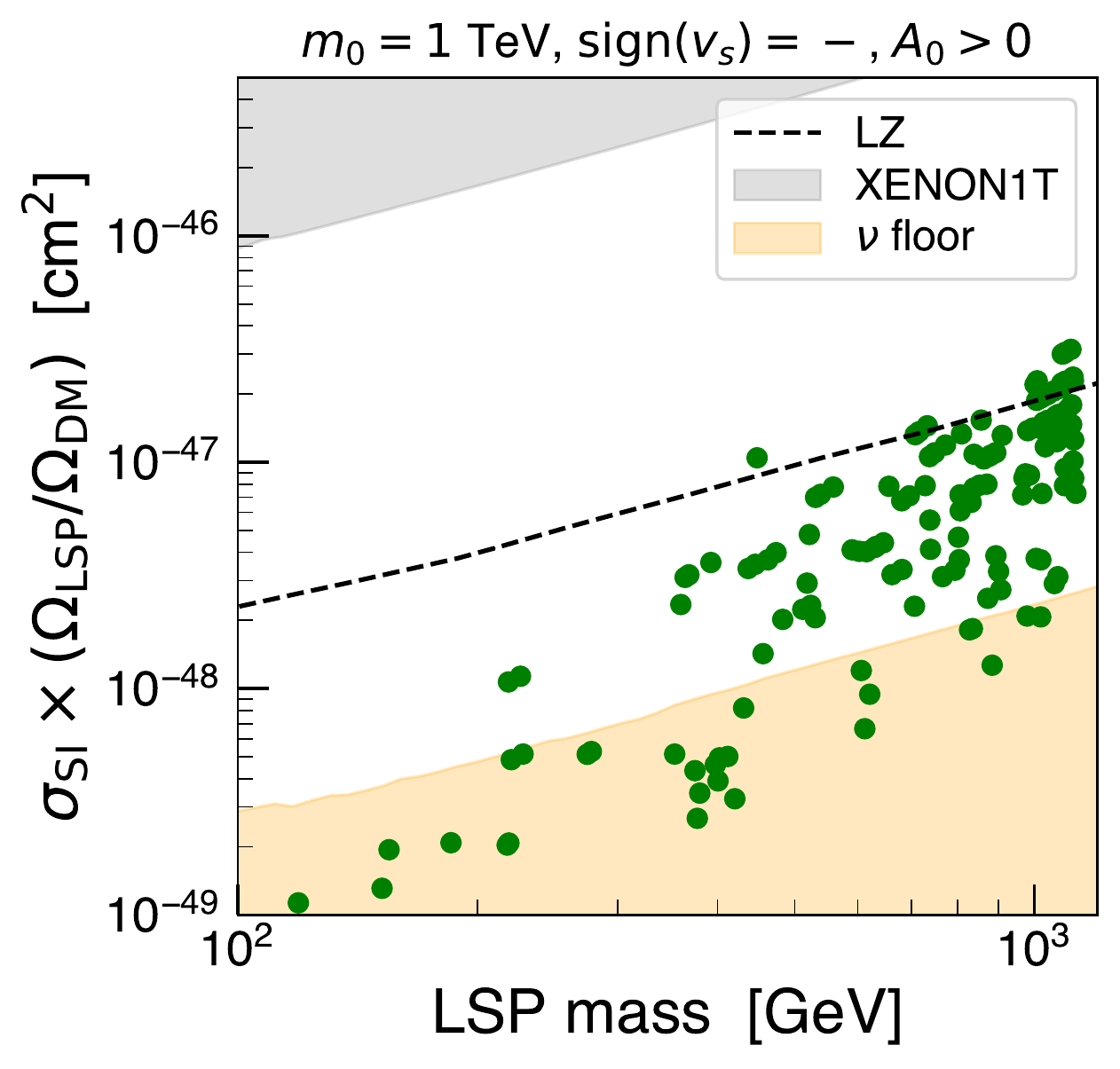}}
\caption{Scatter plots of the spin-independent LSP-proton scattering
 cross sections against the LSP mass.
 Here, the gray shaded region, black
 dashed line, and the orange shaded region represent the XENON1T bound
\cite{Aprile:2018dbl}, the expected sensitivity of the LZ experiment
 \cite{Akerib:2018lyp}, and the neutrino floor \cite{Billard:2013qya},
respectively. 
}
\label{fig:sigsi}
\end{figure}

To study this possibility, in Fig.~\ref{fig:sigsi}, we show the spin-independent
scattering cross section of the LSP with proton, $\sigma_{\text{SI}}$, against the LSP
mass. Here, we only show the parameter points that predict $m_h$ to be in the
favored range and the relic abundance of the LSP 
to be  $\Omega_{\text{LSP}}h^2 \leq \Omega_{\text{DM}}h^2 \simeq
0.12$, where the equality holds for the LSP mass of $\simeq 1$~TeV. 
If $\Omega_{\text{LSP}}h^2 < 0.12$, we expect that the local LSP
density is smaller than the observed local DM density. To take this into
account, we rescale $\sigma_{\text{SI}}$ by a factor of
$\Omega_{\text{LSP}}/ \Omega_{\text{DM}}$ in this plot. The gray shaded region is
excluded by the latest limit imposed by the XENON1T experiment
\cite{Aprile:2018dbl}, while the black dashed line represents the
expected sensitivity of the LZ experiment \cite{Akerib:2018lyp}. The
orange shaded area indicates the neutrino floor \cite{Billard:2013qya},
\textit{i.e.}, the region which the current strategy of the DM direct
searches becomes unable to probe due to the neutrino background. This
figure shows that all of the viable parameter points predict the cross
section well below the current bound. Future multi-ton scale experiments
such as LZ and XENONnT \cite{Aprile:2015uzo} are sensitive to a
part of the model parameter points, and many of the rest are in
principle able to be probed in the future as they are above the neutrino floor.

\section{No-scale/gaugino-mediation type mass spectra}
\label{sec:noscale}

A particularly interesting and more constrained scenario is the case
in which $m_0 = A_0 = 0$ holds at the input scale. This possibility is
motivated by the so-called no-scale \cite{Cremmer:1983bf, Ellis:1983sf,
Ellis:1984bm, Lahanas:1986uc} or gaugino mediation \cite{Kaplan:1999ac,
Chacko:1999mi} models. In the case of the MSSM, it is known that the
no-scale condition $m_0 = A_0 = B_0 = 0$ ($B_0$ denotes the soft
bilinear mass term for the Higgs fields) and the universal gaugino mass
condition $m_{1/2} = M_1 = M_2 = M_3$ at the GUT scale is so restrictive
that viable parameter points cannot be found---it turns out that stau
tends to be either the LSP or tachyonic in most of the parameter
regions \cite{Ellis:2001kg}. To alleviate the problem, one often
takes the input scale to be much higher than the GUT scale, in a
similar manner to the super-GUT models \cite{Ellis:2010ip,
Ellis:2016tjc}, so that the RG effect from the input scale to the GUT
scale generates non-zero soft terms at the GUT scale \cite{Ellis:2001kg, Schmaltz:2000gy,
Schmaltz:2000ei, Ellis:2017djk}. Here, we adopt a different approach to
resolving this problem. We show in what follows that by going to the NMSSM and
allowing the non-universal gaugino masses,\footnote{Gaugino
non-universality in the framework of the gaugino mediation 
in the MSSM is discussed in Refs.~\cite{Yanagida:2013ah,
Yanagida:2013uka}.} we can find viable parameter
points even though all of the soft mass parameters except for the gaugino
masses are set to be zero at the GUT scale.

\begin{table}[t!]
 \begin{center}
\caption{Typical mass spectrum of the NMSSM with non-universal gaugino
  masses and the no-scale condition. }
\label{tab:exnosc}
\vspace{5pt}
\begin{tabular}{ll|ll|ll}
\hline
\hline
\rowcolor{LightGray}
\multicolumn{6}{c}{GUT-scale input parameters ($M_{\text{GUT}} =
 7.34\times 10^{15}$~GeV)} \\
\hline
$m_0$ & $0$ & $A_0$ & $0$ &  &  \\
$M_1$ & $28.30$~TeV & $M_2$ & $16.79$~TeV & $M_3$ & $7.573$~TeV \\
\hline
\rowcolor{LightGray}
\multicolumn{6}{c}{SUSY-scale input parameters} \\
\hline
$\text{sign} (v_s) $ & $-$ & $\lambda$ & $0.255$  &  &  \\
\hline
\rowcolor{LightGray}
\multicolumn{6}{c}{Output parameters} \\
\hline
$\tan\beta$ & $20.03$ &$\kappa$ & $0.412$ & $\mu_{\text{eff}}$ &
		     $-1.031$~TeV \\
$A_\lambda$ & $-3.653$~TeV & $A_\kappa$ & $3.916$~TeV &$m_S^2$ & $-
		     4.851 $~TeV$^2$
\\
$M_1$ & $13.58$~TeV & $M_2$ & $13.91$~TeV & $M_3$ & $14.03$~TeV \\
\hline
\rowcolor{LightGray}
\multicolumn{6}{c}{Mass spectrum} \\
\hline
$m_{h_1}$ & $123.6$~GeV & $m_{h_2}$& $3.190$~TeV &$m_{h_3}$ &
		     $10.59$~TeV \\
$m_{a_1}$ & $1.400$~TeV & $m_{a_2}$& $10.59$~TeV &$m_{H^\pm}$ &
		     $10.59$~TeV \\
$m_{\widetilde{\chi}^0_1}$ & $1.069$~TeV & $m_{\widetilde{\chi}^\pm_1}$&
	     $1.069$~TeV &$m_{\widetilde{g}}$ & $14.69$~TeV \\
$m_{\widetilde{t}_1}$ & $10.14$~TeV & $m_{\widetilde{b}_1}$& $12.01$~TeV
	     &$m_{\widetilde{\tau}_1}$ & $9.905$~TeV \\
\hline
\multicolumn{2}{l}{Other sfermions:} & 
\multicolumn{2}{l|}{$10.2$--$15.6$~TeV} && \\
\hline
\rowcolor{LightGray}
\multicolumn{6}{c}{Dark matter} \\
\hline
$\Omega_{\text{LSP}} h^2$ & $0.103$ & $\sigma_{\text{SI}}^{(p)}$ & $1.8
	     \times 10^{-48}~\text{cm}^2$  & $\sigma_{\text{SD}}^{(p)}$ & $1.8
	     \times 10^{-44}~\text{cm}^2$ \\
\hline
\rowcolor{LightGray}
\multicolumn{6}{c}{Couplings at the GUT scale} \\
\hline
$g_1$ & $0.684$ & $g_2$ & $0.684$ & $g_3$ & $0.686$ \\
$y_t$ & $0.523$ & $y_b$ & $0.121$ & $y_\tau$ & $0.151$ \\
$\lambda$ & $0.287$ & $\kappa$ & $0.561$ & & \\
\hline
\hline
\end{tabular}
 \end{center}
\end{table}

In Table~\ref{tab:exnosc}, we show an example of the phenomenologically
viable mass spectrum of the NMSSM with non-universal gaugino masses and
the no-scale condition $m_0 = A_0 = 0$. As we see, the predicted values
of both $m_h$ and $\Omega_{\text{LSP}} h^2$ are in good agreement with
the observed values. Contrary to the case in Table~\ref{tab:ex1}, in
this case the low-energy value of $A_\kappa$ is predicted to be
positive, as suggested by Eq.~\eqref{eq-AkB} with $A_0 = 0$ and $M_2 >
M_3$. Therefore, $\kappa$ should be positive in order to assure $m_a^2 >
0$. The predicted value of the LSP-nucleon scattering cross sections are
fairly small and all of the colored particles are rather heavy---for
this reason, it is difficult to probe this mass spectrum in the future
experiments.

\begin{figure}[t!]
\centering
\subcaptionbox{\label{fig:noscalelamkap}
$M_2/M_3$ vs $\kappa/\lambda$
}
{\includegraphics[width=0.49\textwidth]{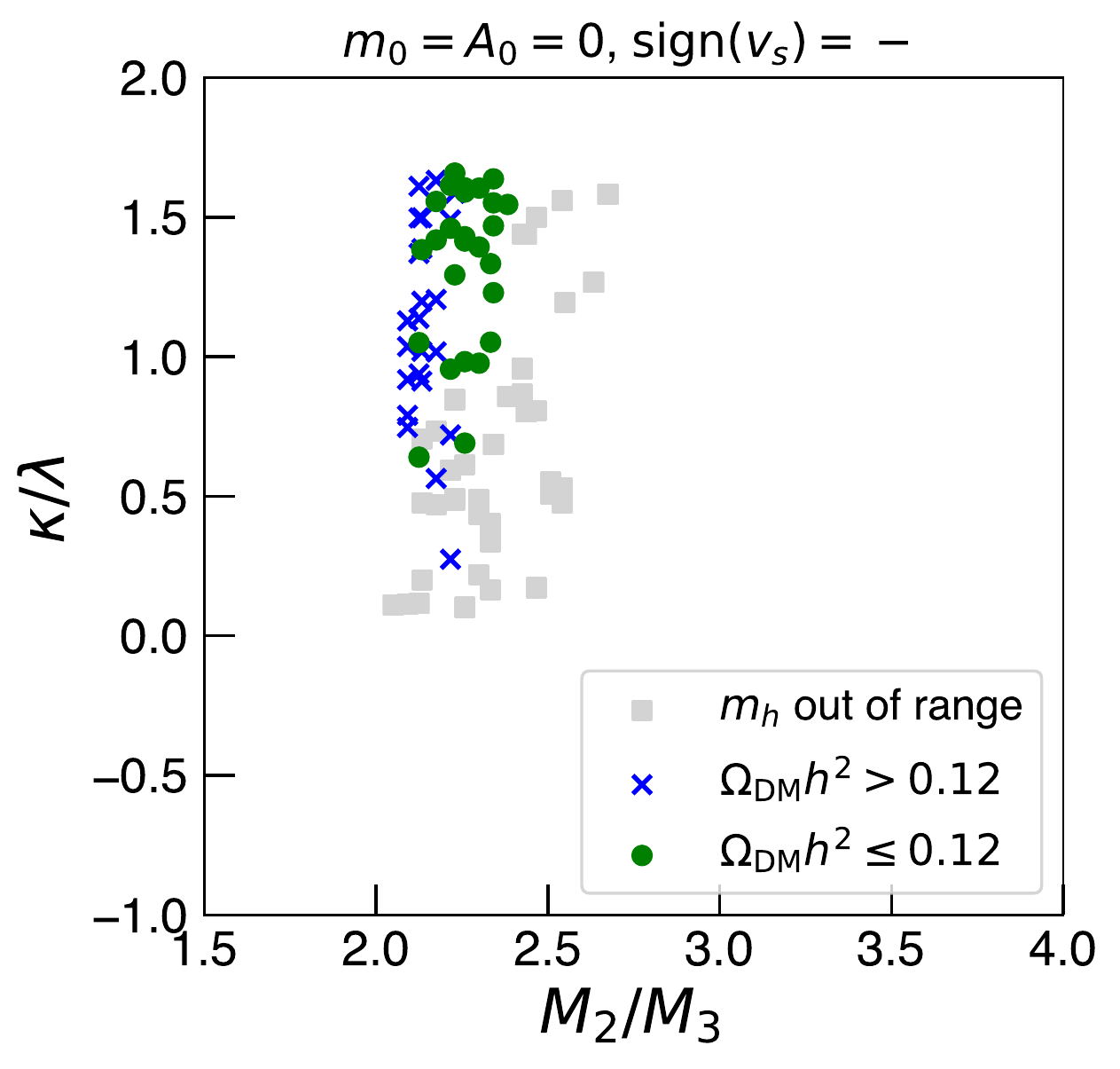}}
\subcaptionbox{\label{fig:noscalemgvsr23}
 $m_{\widetilde{g}}$ vs $M_2/M_3$
}
{\includegraphics[width=0.49\textwidth]{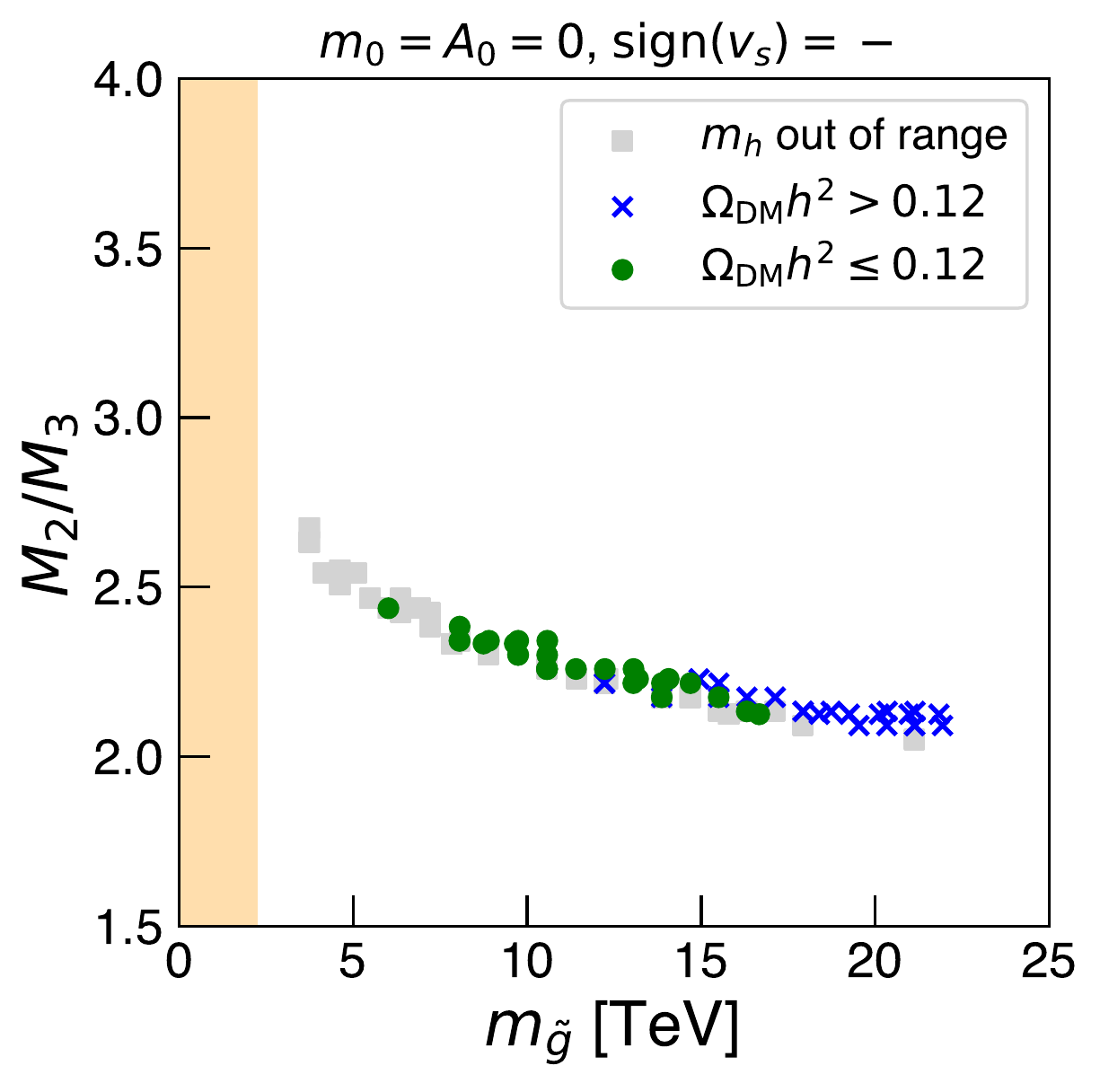}}
\\[5pt]
\subcaptionbox{\label{fig:noscalesigsi}
 LSP mass vs the rescaled $\sigma_{\text{SI}}^{(p)}$
}
{\includegraphics[width=0.49\textwidth]{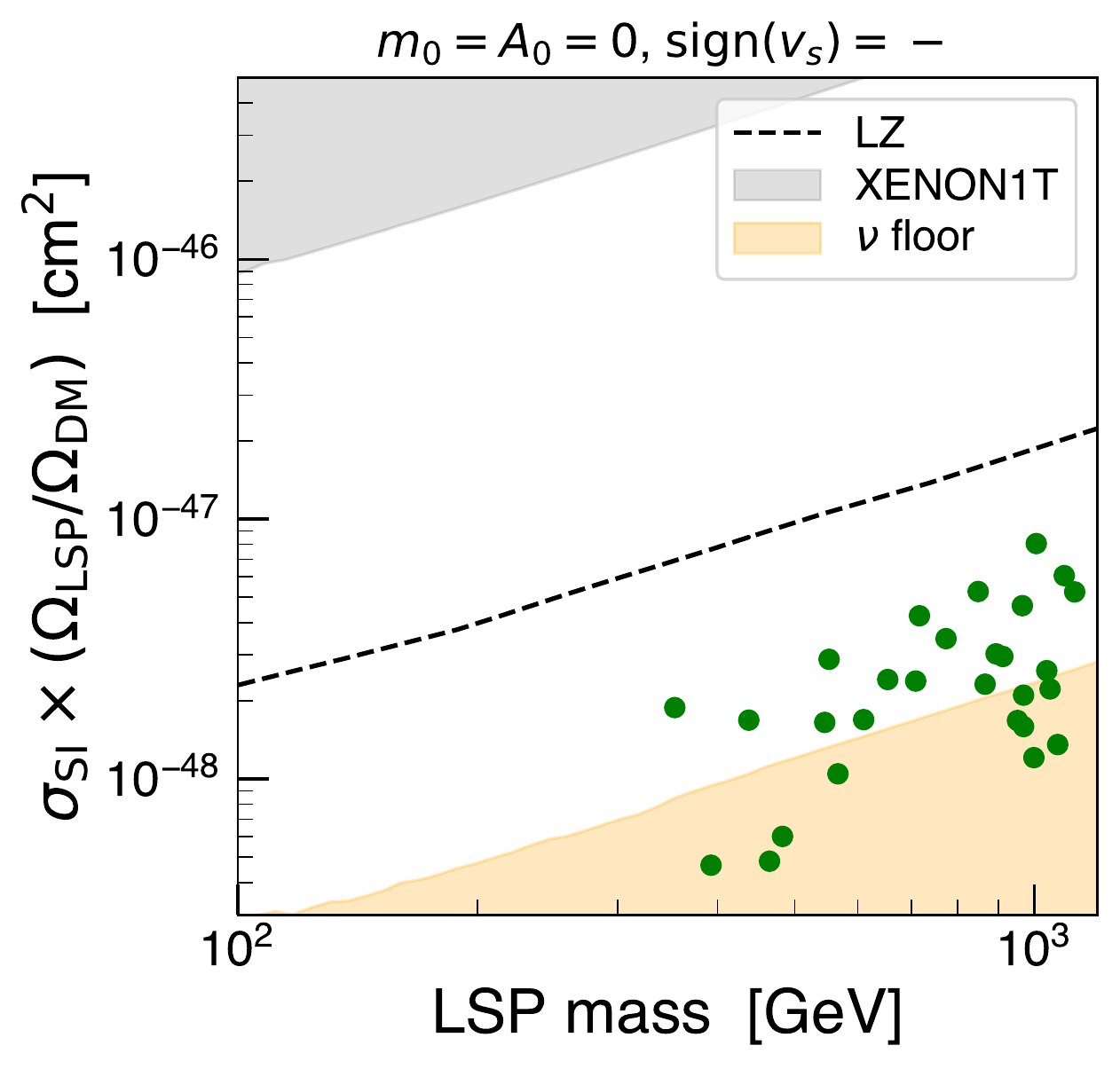}}
\caption{Scatter plots similar to Figs.~\ref{fig:lamkap},
 \ref{fig:mgvsr23}, and \ref{fig:sigsi}.
}
\label{fig:noscale}
\end{figure}

Again, we perform a parameter scan with
\begin{itemize}
 \item $m_0 = A_0 = 0$.
 \item $1~\text{TeV} < M_3 < 12.5$~TeV. 
 \item $1.5 < M_2/M_3 < 4$. 
 \item $0.01 < \lambda < 0.5$.
 \item $\text{sign}(v_s) = -$.
\end{itemize}
The results are summarized in Fig.~\ref{fig:noscale}, where we show
scatter plots similar to Figs.~\ref{fig:lamkap},
\ref{fig:mgvsr23}, and \ref{fig:sigsi} in Figs.~\ref{fig:noscalelamkap},
\ref{fig:noscalemgvsr23}, and \ref{fig:noscalesigsi},
respectively. As we see, the viable parameter points are found only on
the narrow strip lying around $2.0 \lesssim M_2/M_3 \lesssim 2.5$, and the predicted
values of the gluino mass are limited to $5~\text{TeV} \lesssim
m_{\widetilde{g}} \lesssim 17~\text{TeV}$. The SI scattering cross
sections are too small to be probed in the LZ experiment, but may be
reached with a larger detector as quite a few points are above the neutrino
floor.

\section{Summary and Discussion} 
\label{sec:conclusion}

We have discussed the effect of non-universal gaugino masses on the
NMSSM with universal soft trilinear couplings and scalar masses. We have found that if $M_2$ is
large than $M_3$ by a factor of 2--4 at the GUT scale, then constraints
from the tadpole and stability conditions can significantly be
alleviated. Contrary to the CNMSSM, $m_S^2$ runs negative 
at low energies, which can be regarded as the NMSSM counterpart of the
radiative electroweak symmetry breaking in the MSSM \cite{Ibanez:1982fr,
Inoue:1982pi, Ibanez:1982ee, Ellis:1982wr, Ellis:1983bp,
AlvarezGaume:1983gj}.  
The lighter stau, which tends to be either the LSP or
tachyonic in the CNMSSM, can be sufficiently heavy due to the RG effects
from large $M_1$ and $M_2$. We note that such features are generic for
non-universal gaugino masses and thus this type of mass spectrum is
useful to assure the desirable EWSB vacuum even though we slightly relax
the universality conditions of the soft mass parameters. 
In the most of the parameter points, the LSP
is found to be higgsino-like neutralino, and its thermal relic abundance
can account for the observed DM density. We have also shown that many
parameter points in the non-universal gaugino mass scenario can be
tested at a future 100~TeV $pp$ collider or in dark matter direct
searches. 

In our analysis, we have assumed that all of the input parameters are
real just for simplicity. Generically speaking, however, some of these
parameters can be complex, which then introduce additional CP-violating sources
and may generate CP-odd quantities that are experimentally
observable. 
If CP-violating sources exist in the Higgs sector, which can appear at 
tree level~\cite{Ellis:1988er, Matsuda:1995ta, Haba:1996bg, Ham:2001kf, 
Funakubo:2004ka} and/or at loop level~\cite{Garisto:1993ms, Ham:2001wt, 
Ham:2003jf}, there is a mixing between the CP-even and CP-odd Higgs states. 
This mixing changes the properties of the Higgs bosons 
like their masses and couplings to the SM fields. 
An interesting possibility of the CP-violated NMSSM is the electroweak 
baryogenesis (EWBG)~\cite{Kuzmin:1985mm}---in the NMSSM, a strong first-order 
phase transition might have occurred in the early universe, and the extra 
CP-violating sources in the NMSSM scalar potential may bring in the successful 
EWBG~\cite{Pietroni:1992in, Davies:1996qn, Huber:2000mg, Carena:2011jy, 
Cheung:2012pg, Demidov:2016wcv, Akula:2017yfr, Bian:2017wfv}. 
Such CP-violating sources can be probed \cite{King:2015oxa} with 
the measurement of the electric dipole moments (EDMs) of electron, 
neutron, atoms, and so on. 
In the case of the non-universal gaugino mass scenario,
because of the presence of a relatively light higgsino, 
a sizable EDM of electron, $d_e$, may be induced via the Barr-Zee diagrams
\cite{Barr:1990vd}. It is found \cite{Nagata:2014wma, Fukuda:2017jmk}
that $|d_e| \simeq 10^{-(29-30)}$~$e\cdot\text{cm}$ for ${\cal
O}(10)$~TeV gaugino masses if the relative CP phase
between $\mu_{\text{eff}}$ and $M_2$ is sizable. This is below the current bound imposed by
the ACME Collaboration \cite{Baron:2013eja}, $|d_e| < 8.7 \times
10^{-29}~e\cdot \text{cm}$, but can be probe in the future experiments
as their sensitivities are expected to reach $|d_e| \sim 10^{-30}~e\cdot
\text{cm}$ \cite{2012NJPh14j3051K, Kawall:2011zz}.

From the above analyses, we see that the favored values of $M_2/M_3$ are
limited to the range $2\lesssim M_2/M_3 \lesssim 4$. This consequence
may have important implications if we consider a UV completion of this
scenario above the GUT scale. For instance, if we assume a specific GUT
model to realize a non-universal gaugino mass spectrum, the ratio
$M_2/M_3$ is determined to be a particular value \cite{Martin:2009ad},
which then constrains viable model parameter space. Moreover, if we
consider a concrete GUT model, we can also discuss phenomena associated
with the GUT-scale physics, such as proton decay and gauge/Yukawa
coupling unification. As we have seen above, the non-universal gaugino
mass scenario tends to predict a low GUT scale, and thus proton decay
lifetime may be rather short and within the reach of future proton decay
experiments. More detailed discussions on the non-universal gaugino
mass models in the framework of GUTs will be given on another occasion \cite{KKN}.

\section*{Acknowledgments}

This work is supported in part by the Grant-in-Aid for Scientific
Research for Young Scientists (No.18K13534 [JK]), on Innovative Areas
(No.17H05395 [TK], No.18H05542 [NN]), and for Young Scientists B
(No.17K14270 [NN]).

{\small 
\bibliographystyle{JHEP}
\bibliography{ref}

\providecommand{\href}[2]{#2}\begingroup\raggedright\begin{thebibliography}{100}

\bibitem{Kim:1983dt}
J.~E. Kim and H.~P. Nilles, {\it {The mu Problem and the Strong CP Problem}},
  {\em Phys. Lett.} {\bf 138B} (1984) 150--154.

\bibitem{Ellwanger:2009dp}
U.~Ellwanger, C.~Hugonie, and A.~M. Teixeira, {\it {The Next-to-Minimal
  Supersymmetric Standard Model}},  {\em Phys. Rept.} {\bf 496} (2010) 1--77,
  [\href{http://arxiv.org/abs/0910.1785}{{\tt arXiv:0910.1785}}].

\bibitem{Maniatis:2009re}
M.~Maniatis, {\it {The Next-to-Minimal Supersymmetric extension of the Standard
  Model reviewed}},  {\em Int. J. Mod. Phys.} {\bf A25} (2010) 3505--3602,
  [\href{http://arxiv.org/abs/0906.0777}{{\tt arXiv:0906.0777}}].

\bibitem{Ellis:2012aa}
J.~Ellis and K.~A. Olive, {\it {Revisiting the Higgs Mass and Dark Matter in
  the CMSSM}},  {\em Eur. Phys. J.} {\bf C72} (2012) 2005,
  [\href{http://arxiv.org/abs/1202.3262}{{\tt arXiv:1202.3262}}].

\bibitem{Baer:2012uya}
H.~Baer, V.~Barger, and A.~Mustafayev, {\it {Neutralino dark matter in
  mSUGRA/CMSSM with a 125 GeV light Higgs scalar}},  {\em JHEP} {\bf 05} (2012)
  091, [\href{http://arxiv.org/abs/1202.4038}{{\tt arXiv:1202.4038}}].

\bibitem{Baer:2012mv}
H.~Baer, V.~Barger, P.~Huang, D.~Mickelson, A.~Mustafayev, and X.~Tata, {\it
  {Post-LHC7 fine-tuning in the minimal supergravity/CMSSM model with a 125 GeV
  Higgs boson}},  {\em Phys. Rev.} {\bf D87} (2013), no.~3 035017,
  [\href{http://arxiv.org/abs/1210.3019}{{\tt arXiv:1210.3019}}].

\bibitem{Ellis:2012nv}
J.~Ellis, F.~Luo, K.~A. Olive, and P.~Sandick, {\it {The Higgs Mass beyond the
  CMSSM}},  {\em Eur. Phys. J.} {\bf C73} (2013), no.~4 2403,
  [\href{http://arxiv.org/abs/1212.4476}{{\tt arXiv:1212.4476}}].

\bibitem{Liu:2013ula}
M.~Liu and P.~Nath, {\it {Higgs boson mass, proton decay, naturalness, and
  constraints of the LHC and Planck data}},  {\em Phys. Rev.} {\bf D87} (2013),
  no.~9 095012, [\href{http://arxiv.org/abs/1303.7472}{{\tt arXiv:1303.7472}}].

\bibitem{Buchmueller:2013psa}
O.~Buchmueller et~al., {\it {Implications of Improved Higgs Mass Calculations
  for Supersymmetric Models}},  {\em Eur. Phys. J.} {\bf C74} (2014), no.~3
  2809, [\href{http://arxiv.org/abs/1312.5233}{{\tt arXiv:1312.5233}}].

\bibitem{Buchmueller:2013rsa}
O.~Buchmueller et~al., {\it {The CMSSM and NUHM1 after LHC Run 1}},  {\em Eur.
  Phys. J.} {\bf C74} (2014), no.~6 2922,
  [\href{http://arxiv.org/abs/1312.5250}{{\tt arXiv:1312.5250}}].

\bibitem{Roszkowski:2014wqa}
L.~Roszkowski, E.~M. Sessolo, and A.~J. Williams, {\it {What next for the CMSSM
  and the NUHM: Improved prospects for superpartner and dark matter
  detection}},  {\em JHEP} {\bf 08} (2014) 067,
  [\href{http://arxiv.org/abs/1405.4289}{{\tt arXiv:1405.4289}}].

\bibitem{Buchmueller:2015uqa}
O.~Buchmueller, M.~Citron, J.~Ellis, S.~Guha, J.~Marrouche, K.~A. Olive,
  K.~de~Vries, and J.~Zheng, {\it {Collider Interplay for Supersymmetry, Higgs
  and Dark Matter}},  {\em Eur. Phys. J.} {\bf C75} (2015), no.~10 469,
  [\href{http://arxiv.org/abs/1505.04702}{{\tt arXiv:1505.04702}}]. [Erratum:
  Eur. Phys. J.C76,no.4,190(2016)].

\bibitem{Bagnaschi:2015eha}
E.~A. Bagnaschi et~al., {\it {Supersymmetric Dark Matter after LHC Run 1}},
  {\em Eur. Phys. J.} {\bf C75} (2015) 500,
  [\href{http://arxiv.org/abs/1508.01173}{{\tt arXiv:1508.01173}}].

\bibitem{Ellis:2015rya}
J.~Ellis, J.~L. Evans, F.~Luo, N.~Nagata, K.~A. Olive, and P.~Sandick, {\it
  {Beyond the CMSSM without an Accelerator: Proton Decay and Direct Dark Matter
  Detection}},  {\em Eur. Phys. J.} {\bf C76} (2016), no.~1 8,
  [\href{http://arxiv.org/abs/1509.08838}{{\tt arXiv:1509.08838}}].

\bibitem{Athron:2017qdc}
{\bf GAMBIT} Collaboration, P.~Athron et~al., {\it {Global fits of GUT-scale
  SUSY models with GAMBIT}},  {\em Eur. Phys. J.} {\bf C77} (2017), no.~12 824,
  [\href{http://arxiv.org/abs/1705.07935}{{\tt arXiv:1705.07935}}].

\bibitem{Ajaib:2017iyl}
M.~A. Ajaib and I.~Gogoladze, {\it {Status Update on Popular SUSY Models}},
  \href{http://arxiv.org/abs/1710.07842}{{\tt arXiv:1710.07842}}.

\bibitem{Costa:2017gup}
J.~C. Costa et~al., {\it {Likelihood Analysis of the Sub-GUT MSSM in Light of
  LHC 13-TeV Data}},  {\em Eur. Phys. J.} {\bf C78} (2018), no.~2 158,
  [\href{http://arxiv.org/abs/1711.00458}{{\tt arXiv:1711.00458}}].

\bibitem{Ellis:1988er}
J.~R. Ellis, J.~F. Gunion, H.~E. Haber, L.~Roszkowski, and F.~Zwirner, {\it
  {Higgs Bosons in a Nonminimal Supersymmetric Model}},  {\em Phys. Rev.} {\bf
  D39} (1989) 844.

\bibitem{Ellwanger:1993xa}
U.~Ellwanger, M.~Rausch~de Traubenberg, and C.~A. Savoy, {\it {Particle
  spectrum in supersymmetric models with a gauge singlet}},  {\em Phys. Lett.}
  {\bf B315} (1993) 331--337, [\href{http://arxiv.org/abs/hep-ph/9307322}{{\tt
  hep-ph/9307322}}].

\bibitem{Elliott:1994ht}
T.~Elliott, S.~F. King, and P.~L. White, {\it {Unification constraints in the
  next-to-minimal supersymmetric standard model}},  {\em Phys. Lett.} {\bf
  B351} (1995) 213--219, [\href{http://arxiv.org/abs/hep-ph/9406303}{{\tt
  hep-ph/9406303}}].

\bibitem{Ellwanger:1995ru}
U.~Ellwanger, M.~Rausch~de Traubenberg, and C.~A. Savoy, {\it {Higgs
  phenomenology of the supersymmetric model with a gauge singlet}},  {\em Z.
  Phys.} {\bf C67} (1995) 665--670,
  [\href{http://arxiv.org/abs/hep-ph/9502206}{{\tt hep-ph/9502206}}].

\bibitem{King:1995vk}
S.~F. King and P.~L. White, {\it {Resolving the constrained minimal and
  next-to-minimal supersymmetric standard models}},  {\em Phys. Rev.} {\bf D52}
  (1995) 4183--4216, [\href{http://arxiv.org/abs/hep-ph/9505326}{{\tt
  hep-ph/9505326}}].

\bibitem{Ellwanger:1996gw}
U.~Ellwanger, M.~Rausch~de Traubenberg, and C.~A. Savoy, {\it {Phenomenology of
  supersymmetric models with a singlet}},  {\em Nucl. Phys.} {\bf B492} (1997)
  21--50, [\href{http://arxiv.org/abs/hep-ph/9611251}{{\tt hep-ph/9611251}}].

\bibitem{Djouadi:2008yj}
A.~Djouadi, U.~Ellwanger, and A.~M. Teixeira, {\it {The Constrained
  next-to-minimal supersymmetric standard model}},  {\em Phys. Rev. Lett.} {\bf
  101} (2008) 101802, [\href{http://arxiv.org/abs/0803.0253}{{\tt
  arXiv:0803.0253}}].

\bibitem{Djouadi:2008uj}
A.~Djouadi, U.~Ellwanger, and A.~M. Teixeira, {\it {Phenomenology of the
  constrained NMSSM}},  {\em JHEP} {\bf 04} (2009) 031,
  [\href{http://arxiv.org/abs/0811.2699}{{\tt arXiv:0811.2699}}].

\bibitem{Arbey:2011ab}
A.~Arbey, M.~Battaglia, A.~Djouadi, F.~Mahmoudi, and J.~Quevillon, {\it
  {Implications of a 125 GeV Higgs for supersymmetric models}},  {\em Phys.
  Lett.} {\bf B708} (2012) 162--169,
  [\href{http://arxiv.org/abs/1112.3028}{{\tt arXiv:1112.3028}}].

\bibitem{Aad:2015zhl}
{\bf ATLAS, CMS} Collaboration, G.~Aad et~al., {\it {Combined Measurement of
  the Higgs Boson Mass in $pp$ Collisions at $\sqrt{s}=7$ and 8 TeV with the
  ATLAS and CMS Experiments}},  {\em Phys. Rev. Lett.} {\bf 114} (2015) 191803,
  [\href{http://arxiv.org/abs/1503.07589}{{\tt arXiv:1503.07589}}].

\bibitem{Aghanim:2018eyx}
{\bf Planck} Collaboration, N.~Aghanim et~al., {\it {Planck 2018 results. VI.
  Cosmological parameters}},  \href{http://arxiv.org/abs/1807.06209}{{\tt
  arXiv:1807.06209}}.

\bibitem{Ellwanger:2006rn}
U.~Ellwanger and C.~Hugonie, {\it {NMSPEC: A Fortran code for the sparticle and
  Higgs masses in the NMSSM with GUT scale boundary conditions}},  {\em Comput.
  Phys. Commun.} {\bf 177} (2007) 399--407,
  [\href{http://arxiv.org/abs/hep-ph/0612134}{{\tt hep-ph/0612134}}].

\bibitem{Gunion:2012zd}
J.~F. Gunion, Y.~Jiang, and S.~Kraml, {\it {The Constrained NMSSM and Higgs
  near 125 GeV}},  {\em Phys. Lett.} {\bf B710} (2012) 454--459,
  [\href{http://arxiv.org/abs/1201.0982}{{\tt arXiv:1201.0982}}].

\bibitem{Ellwanger:2012ke}
U.~Ellwanger and C.~Hugonie, {\it {Higgs bosons near 125 GeV in the NMSSM with
  constraints at the GUT scale}},  {\em Adv. High Energy Phys.} {\bf 2012}
  (2012) 625389, [\href{http://arxiv.org/abs/1203.5048}{{\tt
  arXiv:1203.5048}}].

\bibitem{Kowalska:2012gs}
K.~Kowalska, S.~Munir, L.~Roszkowski, E.~M. Sessolo, S.~Trojanowski, and
  Y.-L.~S. Tsai, {\it {Constrained next-to-minimal supersymmetric standard
  model with a 126 GeV Higgs boson: A global analysis}},  {\em Phys. Rev.} {\bf
  D87} (2013) 115010, [\href{http://arxiv.org/abs/1211.1693}{{\tt
  arXiv:1211.1693}}].

\bibitem{Das:2013ta}
D.~Das, U.~Ellwanger, and A.~M. Teixeira, {\it {LHC constraints on $M_{1/2}$
  and $m_0$ in the semi-constrained NMSSM}},  {\em JHEP} {\bf 04} (2013) 117,
  [\href{http://arxiv.org/abs/1301.7584}{{\tt arXiv:1301.7584}}].

\bibitem{Beskidt:2013gia}
C.~Beskidt, W.~de~Boer, and D.~I. Kazakov, {\it {A comparison of the Higgs
  sectors of the CMSSM and NMSSM for a 126 GeV Higgs boson}},  {\em Phys.
  Lett.} {\bf B726} (2013) 758--766,
  [\href{http://arxiv.org/abs/1308.1333}{{\tt arXiv:1308.1333}}].

\bibitem{Ellwanger:2014dfa}
U.~Ellwanger and C.~Hugonie, {\it {The semi-constrained NMSSM satisfying bounds
  from the LHC, LUX and Planck}},  {\em JHEP} {\bf 08} (2014) 046,
  [\href{http://arxiv.org/abs/1405.6647}{{\tt arXiv:1405.6647}}].

\bibitem{Beskidt:2016egy}
C.~Beskidt, W.~de~Boer, D.~I. Kazakov, and S.~Wayand, {\it {Higgs branching
  ratios in constrained minimal and next-to-minimal supersymmetry scenarios
  surveyed}},  {\em Phys. Lett.} {\bf B759} (2016) 141--148,
  [\href{http://arxiv.org/abs/1602.08707}{{\tt arXiv:1602.08707}}].

\bibitem{Cerdeno:2017sks}
D.~G. Cerde\~{n}o, V.~De~Romeri, V.~Mart\'{i}n-Lozano, K.~A. Olive, and
  O.~Seto, {\it {The Constrained NMSSM with right-handed neutrinos}},  {\em
  Eur. Phys. J.} {\bf C78} (2018), no.~4 290,
  [\href{http://arxiv.org/abs/1707.03990}{{\tt arXiv:1707.03990}}].

\bibitem{Choi:2004sx}
K.~Choi, A.~Falkowski, H.~P. Nilles, M.~Olechowski, and S.~Pokorski, {\it
  {Stability of flux compactifications and the pattern of supersymmetry
  breaking}},  {\em JHEP} {\bf 11} (2004) 076,
  [\href{http://arxiv.org/abs/hep-th/0411066}{{\tt hep-th/0411066}}].

\bibitem{Choi:2005ge}
K.~Choi, A.~Falkowski, H.~P. Nilles, and M.~Olechowski, {\it {Soft
  supersymmetry breaking in KKLT flux compactification}},  {\em Nucl. Phys.}
  {\bf B718} (2005) 113--133, [\href{http://arxiv.org/abs/hep-th/0503216}{{\tt
  hep-th/0503216}}].

\bibitem{Endo:2005uy}
M.~Endo, M.~Yamaguchi, and K.~Yoshioka, {\it {A Bottom-up approach to moduli
  dynamics in heavy gravitino scenario: Superpotential, soft terms and
  sparticle mass spectrum}},  {\em Phys. Rev.} {\bf D72} (2005) 015004,
  [\href{http://arxiv.org/abs/hep-ph/0504036}{{\tt hep-ph/0504036}}].

\bibitem{Choi:2005uz}
K.~Choi, K.~S. Jeong, and K.-i. Okumura, {\it {Phenomenology of mixed
  modulus-anomaly mediation in fluxed string compactifications and brane
  models}},  {\em JHEP} {\bf 09} (2005) 039,
  [\href{http://arxiv.org/abs/hep-ph/0504037}{{\tt hep-ph/0504037}}].

\bibitem{Choi:2005hd}
K.~Choi, K.~S. Jeong, T.~Kobayashi, and K.-i. Okumura, {\it {Little SUSY
  hierarchy in mixed modulus-anomaly mediation}},  {\em Phys. Lett.} {\bf B633}
  (2006) 355--361, [\href{http://arxiv.org/abs/hep-ph/0508029}{{\tt
  hep-ph/0508029}}].

\bibitem{Kitano:2005wc}
R.~Kitano and Y.~Nomura, {\it {A Solution to the supersymmetric fine-tuning
  problem within the MSSM}},  {\em Phys. Lett.} {\bf B631} (2005) 58--67,
  [\href{http://arxiv.org/abs/hep-ph/0509039}{{\tt hep-ph/0509039}}].

\bibitem{Choi:2006xb}
K.~Choi, K.~S. Jeong, T.~Kobayashi, and K.-i. Okumura, {\it {TeV Scale Mirage
  Mediation and Natural Little SUSY Hierarchy}},  {\em Phys. Rev.} {\bf D75}
  (2007) 095012, [\href{http://arxiv.org/abs/hep-ph/0612258}{{\tt
  hep-ph/0612258}}].

\bibitem{Ellis:1984bm}
J.~R. Ellis, C.~Kounnas, and D.~V. Nanopoulos, {\it {No Scale Supersymmetric
  Guts}},  {\em Nucl. Phys.} {\bf B247} (1984) 373--395.

\bibitem{Ellis:1985jn}
J.~R. Ellis, K.~Enqvist, D.~V. Nanopoulos, and K.~Tamvakis, {\it {Gaugino
  Masses and Grand Unification}},  {\em Phys. Lett.} {\bf 155B} (1985)
  381--386.

\bibitem{Drees:1985bx}
M.~Drees, {\it {Phenomenological Consequences of $N=1$ Supergravity Theories
  With Nonminimal Kinetic Energy Terms for Vector Superfields}},  {\em Phys.
  Lett.} {\bf 158B} (1985) 409--412.

\bibitem{Anderson:1996bg}
G.~Anderson, C.~H. Chen, J.~F. Gunion, J.~D. Lykken, T.~Moroi, and Y.~Yamada,
  {\it {Motivations for and implications of nonuniversal GUT scale boundary
  conditions for soft SUSY breaking parameters}},  {\em eConf} {\bf C960625}
  (1996) SUP107, [\href{http://arxiv.org/abs/hep-ph/9609457}{{\tt
  hep-ph/9609457}}]. [,669(1996)].

\bibitem{Chakrabortty:2008zk}
J.~Chakrabortty and A.~Raychaudhuri, {\it {A Note on dimension-5 operators in
  GUTs and their impact}},  {\em Phys. Lett.} {\bf B673} (2009) 57--62,
  [\href{http://arxiv.org/abs/0812.2783}{{\tt arXiv:0812.2783}}].

\bibitem{Martin:2009ad}
S.~P. Martin, {\it {Non-universal gaugino masses from non-singlet F-terms in
  non-minimal unified models}},  {\em Phys. Rev.} {\bf D79} (2009) 095019,
  [\href{http://arxiv.org/abs/0903.3568}{{\tt arXiv:0903.3568}}].

\bibitem{Chakrabortty:2010xq}
J.~Chakrabortty and A.~Raychaudhuri, {\it {Dimension-5 operators and the
  unification condition in SO(10) and E(6)}},
  \href{http://arxiv.org/abs/1006.1252}{{\tt arXiv:1006.1252}}.

\bibitem{Younkin:2012ui}
J.~E. Younkin and S.~P. Martin, {\it {Non-universal gaugino masses, the
  supersymmetric little hierarchy problem, and dark matter}},  {\em Phys. Rev.}
  {\bf D85} (2012) 055028, [\href{http://arxiv.org/abs/1201.2989}{{\tt
  arXiv:1201.2989}}].

\bibitem{Kobayashi:2017fgl}
T.~Kobayashi, Y.~Omura, O.~Seto, and K.~Ueda, {\it {Realization of a
  spontaneous gauge and supersymmetry breaking vacuum}},  {\em JHEP} {\bf 11}
  (2017) 073, [\href{http://arxiv.org/abs/1705.00809}{{\tt arXiv:1705.00809}}].

\bibitem{Blumenhagen:2006ci}
R.~Blumenhagen, B.~Kors, D.~Lust, and S.~Stieberger, {\it {Four-dimensional
  String Compactifications with D-Branes, Orientifolds and Fluxes}},  {\em
  Phys. Rept.} {\bf 445} (2007) 1--193,
  [\href{http://arxiv.org/abs/hep-th/0610327}{{\tt hep-th/0610327}}].

\bibitem{Ibanez:2012zz}
L.~E. Ibanez and A.~M. Uranga, {\em {String theory and particle physics: An
  introduction to string phenomenology}}.
\newblock Cambridge University Press, 2012.

\bibitem{Abe:2007kf}
H.~Abe, T.~Kobayashi, and Y.~Omura, {\it {Relaxed fine-tuning in models with
  non-universal gaugino masses}},  {\em Phys. Rev.} {\bf D76} (2007) 015002,
  [\href{http://arxiv.org/abs/hep-ph/0703044}{{\tt hep-ph/0703044}}].

\bibitem{Abe:2012xm}
H.~Abe, J.~Kawamura, and H.~Otsuka, {\it {The Higgs boson mass in a natural
  MSSM with nonuniversal gaugino masses at the GUT scale}},  {\em PTEP} {\bf
  2013} (2013) 013B02, [\href{http://arxiv.org/abs/1208.5328}{{\tt
  arXiv:1208.5328}}].

\bibitem{Bhattacharya:2007dr}
S.~Bhattacharya, A.~Datta, and B.~Mukhopadhyaya, {\it {Non-universal gaugino
  masses: A Signal-based analysis for the Large Hadron Collider}},  {\em JHEP}
  {\bf 10} (2007) 080, [\href{http://arxiv.org/abs/0708.2427}{{\tt
  arXiv:0708.2427}}].

\bibitem{Bhattacharya:2009wv}
S.~Bhattacharya and J.~Chakrabortty, {\it {Gaugino mass non-universality in an
  SO(10) supersymmetric Grand Unified Theory: Low-energy spectra and collider
  signals}},  {\em Phys. Rev.} {\bf D81} (2010) 015007,
  [\href{http://arxiv.org/abs/0903.4196}{{\tt arXiv:0903.4196}}].

\bibitem{Horton:2009ed}
D.~Horton and G.~G. Ross, {\it {Naturalness and Focus Points with Non-Universal
  Gaugino Masses}},  {\em Nucl. Phys.} {\bf B830} (2010) 221--247,
  [\href{http://arxiv.org/abs/0908.0857}{{\tt arXiv:0908.0857}}].

\bibitem{Brummer:2012zc}
F.~Brummer and W.~Buchmuller, {\it {The Fermi scale as a focus point of
  high-scale gauge mediation}},  {\em JHEP} {\bf 05} (2012) 006,
  [\href{http://arxiv.org/abs/1201.4338}{{\tt arXiv:1201.4338}}].

\bibitem{Gogoladze:2012yf}
I.~Gogoladze, F.~Nasir, and Q.~Shafi, {\it {Non-Universal Gaugino Masses and
  Natural Supersymmetry}},  {\em Int. J. Mod. Phys.} {\bf A28} (2013) 1350046,
  [\href{http://arxiv.org/abs/1212.2593}{{\tt arXiv:1212.2593}}].

\bibitem{Yanagida:2013ah}
T.~T. Yanagida and N.~Yokozaki, {\it {Focus Point in Gaugino Mediation ~
  Reconsideration of the Fine-tuning Problem ~}},  {\em Phys. Lett.} {\bf B722}
  (2013) 355--359, [\href{http://arxiv.org/abs/1301.1137}{{\tt
  arXiv:1301.1137}}].

\bibitem{Gogoladze:2013wva}
I.~Gogoladze, F.~Nasir, and Q.~Shafi, {\it {SO(10) as a Framework for Natural
  Supersymmetry}},  {\em JHEP} {\bf 11} (2013) 173,
  [\href{http://arxiv.org/abs/1306.5699}{{\tt arXiv:1306.5699}}].

\bibitem{Yanagida:2013uka}
T.~T. Yanagida and N.~Yokozaki, {\it {Bino-Higgsino Mixed Dark Matter in a
  Focus Point Gaugino Mediation}},  {\em JHEP} {\bf 11} (2013) 020,
  [\href{http://arxiv.org/abs/1308.0536}{{\tt arXiv:1308.0536}}].

\bibitem{Chakrabortty:2013voa}
J.~Chakrabortty, S.~Mohanty, and S.~Rao, {\it {Non-universal gaugino mass GUT
  models in the light of dark matter and LHC constraints}},  {\em JHEP} {\bf
  02} (2014) 074, [\href{http://arxiv.org/abs/1310.3620}{{\tt
  arXiv:1310.3620}}].

\bibitem{Martin:2013aha}
S.~P. Martin, {\it {Nonuniversal gaugino masses and seminatural supersymmetry
  in view of the Higgs boson discovery}},  {\em Phys. Rev.} {\bf D89} (2014),
  no.~3 035011, [\href{http://arxiv.org/abs/1312.0582}{{\tt arXiv:1312.0582}}].

\bibitem{Chakrabortty:2015ika}
J.~Chakrabortty, A.~Choudhury, and S.~Mondal, {\it {Non-universal Gaugino mass
  models under the lamppost of muon (g-2)}},  {\em JHEP} {\bf 07} (2015) 038,
  [\href{http://arxiv.org/abs/1503.08703}{{\tt arXiv:1503.08703}}].

\bibitem{Harigaya:2015jba}
K.~Harigaya, T.~T. Yanagida, and N.~Yokozaki, {\it {Muon $g-2$ in focus point
  SUSY}},  {\em Phys. Rev.} {\bf D92} (2015), no.~3 035011,
  [\href{http://arxiv.org/abs/1505.01987}{{\tt arXiv:1505.01987}}].

\bibitem{Abe:2015xva}
H.~Abe, J.~Kawamura, and Y.~Omura, {\it {LHC phenomenology of natural MSSM with
  non-universal gaugino masses at the unification scale}},  {\em JHEP} {\bf 08}
  (2015) 089, [\href{http://arxiv.org/abs/1505.03729}{{\tt arXiv:1505.03729}}].

\bibitem{Sumita:2015tba}
K.~Sumita, {\it {Nonuniversal gaugino masses in a magnetized toroidal
  compactification of SYM theories}},  {\em JHEP} {\bf 10} (2015) 156,
  [\href{http://arxiv.org/abs/1507.04408}{{\tt arXiv:1507.04408}}].

\bibitem{Kawamura:2016drh}
J.~Kawamura and Y.~Omura, {\it {Constraints on nonuniversal gaugino mass
  scenario using the latest LHC data}},  {\em Phys. Rev.} {\bf D93} (2016),
  no.~5 055019, [\href{http://arxiv.org/abs/1601.03484}{{\tt
  arXiv:1601.03484}}].

\bibitem{Kawamura:2017amp}
J.~Kawamura and Y.~Omura, {\it {Study of dark matter physics in non-universal
  gaugino mass scenario}},  {\em JHEP} {\bf 08} (2017) 072,
  [\href{http://arxiv.org/abs/1703.10379}{{\tt arXiv:1703.10379}}].

\bibitem{Kawamura:2017qey}
J.~Kawamura and Y.~Omura, {\it {Analysis of the TeV-scale mirage mediation with
  heavy superparticles}},  {\em JHEP} {\bf 11} (2017) 189,
  [\href{http://arxiv.org/abs/1710.03412}{{\tt arXiv:1710.03412}}].

\bibitem{Martin:2017vlf}
S.~P. Martin, {\it {Quasifixed points from scalar sequestering and the little
  hierarchy problem in supersymmetry}},  {\em Phys. Rev.} {\bf D97} (2018),
  no.~3 035006, [\href{http://arxiv.org/abs/1712.05806}{{\tt
  arXiv:1712.05806}}].

\bibitem{Kobayashi:2012ee}
T.~Kobayashi, H.~Makino, K.-i. Okumura, T.~Shimomura, and T.~Takahashi, {\it
  {TeV scale mirage mediation in NMSSM}},  {\em JHEP} {\bf 01} (2013) 081,
  [\href{http://arxiv.org/abs/1204.3561}{{\tt arXiv:1204.3561}}].

\bibitem{Asano:2012sv}
M.~Asano and T.~Higaki, {\it {Natural supersymmetric spectrum in mirage
  mediation}},  {\em Phys. Rev.} {\bf D86} (2012) 035020,
  [\href{http://arxiv.org/abs/1204.0508}{{\tt arXiv:1204.0508}}].

\bibitem{Hagimoto:2015tua}
K.~Hagimoto, T.~Kobayashi, H.~Makino, K.-i. Okumura, and T.~Shimomura, {\it
  {Phenomenology of NMSSM in TeV scale mirage mediation}},  {\em JHEP} {\bf 02}
  (2016) 089, [\href{http://arxiv.org/abs/1509.05327}{{\tt arXiv:1509.05327}}].

\bibitem{Cohen:2013xda}
T.~Cohen, T.~Golling, M.~Hance, A.~Henrichs, K.~Howe, J.~Loyal, S.~Padhi, and
  J.~G. Wacker, {\it {SUSY Simplified Models at 14, 33, and 100 TeV Proton
  Colliders}},  {\em JHEP} {\bf 04} (2014) 117,
  [\href{http://arxiv.org/abs/1311.6480}{{\tt arXiv:1311.6480}}].

\bibitem{Ellis:2015xba}
S.~A.~R. Ellis and B.~Zheng, {\it {Reaching for squarks and gauginos at a 100
  TeV p-p collider}},  {\em Phys. Rev.} {\bf D92} (2015), no.~7 075034,
  [\href{http://arxiv.org/abs/1506.02644}{{\tt arXiv:1506.02644}}].

\bibitem{Arkani-Hamed:2015vfh}
N.~Arkani-Hamed, T.~Han, M.~Mangano, and L.-T. Wang, {\it {Physics
  opportunities of a 100 TeV proton-proton collider}},  {\em Phys. Rept.} {\bf
  652} (2016) 1--49, [\href{http://arxiv.org/abs/1511.06495}{{\tt
  arXiv:1511.06495}}].

\bibitem{Golling:2016gvc}
T.~Golling et~al., {\it {Physics at a 100 TeV pp collider: beyond the Standard
  Model phenomena}},  {\em CERN Yellow Report} (2017), no.~3 441--634,
  [\href{http://arxiv.org/abs/1606.00947}{{\tt arXiv:1606.00947}}].

\bibitem{Pandita:1993tg}
P.~N. Pandita, {\it {Radiative corrections to the scalar Higgs masses in a
  nonminimal supersymmetric Standard Model}},  {\em Z. Phys.} {\bf C59} (1993)
  575--584.

\bibitem{Elliott:1993bs}
T.~Elliott, S.~F. King, and P.~L. White, {\it {Radiative corrections to Higgs
  boson masses in the next-to-minimal supersymmetric Standard Model}},  {\em
  Phys. Rev.} {\bf D49} (1994) 2435--2456,
  [\href{http://arxiv.org/abs/hep-ph/9308309}{{\tt hep-ph/9308309}}].

\bibitem{Ellwanger:1999bv}
U.~Ellwanger and C.~Hugonie, {\it {Constraints from charge and color breaking
  minima in the (M+1)SSM}},  {\em Phys. Lett.} {\bf B457} (1999) 299--306,
  [\href{http://arxiv.org/abs/hep-ph/9902401}{{\tt hep-ph/9902401}}].

\bibitem{Miller:2003ay}
D.~J. Miller, R.~Nevzorov, and P.~M. Zerwas, {\it {The Higgs sector of the
  next-to-minimal supersymmetric standard model}},  {\em Nucl. Phys.} {\bf
  B681} (2004) 3--30, [\href{http://arxiv.org/abs/hep-ph/0304049}{{\tt
  hep-ph/0304049}}].

\bibitem{Funakubo:2004ka}
K.~Funakubo and S.~Tao, {\it {The Higgs sector in the next-to-MSSM}},  {\em
  Prog. Theor. Phys.} {\bf 113} (2005) 821--842,
  [\href{http://arxiv.org/abs/hep-ph/0409294}{{\tt hep-ph/0409294}}].

\bibitem{Cheung:2010ba}
K.~Cheung, T.-J. Hou, J.~S. Lee, and E.~Senaha, {\it {The Higgs Boson Sector of
  the Next-to-MSSM with CP Violation}},  {\em Phys. Rev.} {\bf D82} (2010)
  075007, [\href{http://arxiv.org/abs/1006.1458}{{\tt arXiv:1006.1458}}].

\bibitem{Kanehata:2011ei}
Y.~Kanehata, T.~Kobayashi, Y.~Konishi, O.~Seto, and T.~Shimomura, {\it
  {Constraints from Unrealistic Vacua in the Next-to-Minimal Supersymmetric
  Standard Model}},  {\em Prog. Theor. Phys.} {\bf 126} (2011) 1051--1076,
  [\href{http://arxiv.org/abs/1103.5109}{{\tt arXiv:1103.5109}}].

\bibitem{Kobayashi:2012xv}
T.~Kobayashi, T.~Shimomura, and T.~Takahashi, {\it {Constraining the Higgs
  sector from False Vacua in the Next-to-Minimal Supersymmetric Standard
  Model}},  {\em Phys. Rev.} {\bf D86} (2012) 015029,
  [\href{http://arxiv.org/abs/1203.4328}{{\tt arXiv:1203.4328}}].

\bibitem{Beuria:2016cdk}
J.~Beuria, U.~Chattopadhyay, A.~Datta, and A.~Dey, {\it {Exploring viable vacua
  of the Z$_{3}$-symmetric NMSSM}},  {\em JHEP} {\bf 04} (2017) 024,
  [\href{http://arxiv.org/abs/1612.06803}{{\tt arXiv:1612.06803}}].

\bibitem{Zeldovich:1974uw}
{\relax Ya}.~B. Zeldovich, I.~{\relax Yu}. Kobzarev, and L.~B. Okun, {\it
  {Cosmological Consequences of the Spontaneous Breakdown of Discrete
  Symmetry}},  {\em Zh. Eksp. Teor. Fiz.} {\bf 67} (1974) 3--11. [Sov. Phys.
  JETP40,1(1974)].

\bibitem{Vilenkin:1984ib}
A.~Vilenkin, {\it {Cosmic Strings and Domain Walls}},  {\em Phys. Rept.} {\bf
  121} (1985) 263--315.

\bibitem{Ellis:1986mq}
J.~R. Ellis, K.~Enqvist, D.~V. Nanopoulos, K.~A. Olive, M.~Quiros, and
  F.~Zwirner, {\it {Problems for (2,0) Compactifications}},  {\em Phys. Lett.}
  {\bf B176} (1986) 403--408.

\bibitem{Abel:1995wk}
S.~A. Abel, S.~Sarkar, and P.~L. White, {\it {On the cosmological domain wall
  problem for the minimally extended supersymmetric standard model}},  {\em
  Nucl. Phys.} {\bf B454} (1995) 663--684,
  [\href{http://arxiv.org/abs/hep-ph/9506359}{{\tt hep-ph/9506359}}].

\bibitem{Abel:1996cr}
S.~A. Abel, {\it {Destabilizing divergences in the NMSSM}},  {\em Nucl. Phys.}
  {\bf B480} (1996) 55--72, [\href{http://arxiv.org/abs/hep-ph/9609323}{{\tt
  hep-ph/9609323}}].

\bibitem{Panagiotakopoulos:1998yw}
C.~Panagiotakopoulos and K.~Tamvakis, {\it {Stabilized NMSSM without domain
  walls}},  {\em Phys. Lett.} {\bf B446} (1999) 224--227,
  [\href{http://arxiv.org/abs/hep-ph/9809475}{{\tt hep-ph/9809475}}].

\bibitem{Panagiotakopoulos:1999ah}
C.~Panagiotakopoulos and K.~Tamvakis, {\it {New minimal extension of MSSM}},
  {\em Phys. Lett.} {\bf B469} (1999) 145--148,
  [\href{http://arxiv.org/abs/hep-ph/9908351}{{\tt hep-ph/9908351}}].

\bibitem{Panagiotakopoulos:2000wp}
C.~Panagiotakopoulos and A.~Pilaftsis, {\it {Higgs scalars in the minimal
  nonminimal supersymmetric standard model}},  {\em Phys. Rev.} {\bf D63}
  (2001) 055003, [\href{http://arxiv.org/abs/hep-ph/0008268}{{\tt
  hep-ph/0008268}}].

\bibitem{Dedes:2000jp}
A.~Dedes, C.~Hugonie, S.~Moretti, and K.~Tamvakis, {\it {Phenomenology of a new
  minimal supersymmetric extension of the standard model}},  {\em Phys. Rev.}
  {\bf D63} (2001) 055009, [\href{http://arxiv.org/abs/hep-ph/0009125}{{\tt
  hep-ph/0009125}}].

\bibitem{Hamaguchi:2011nm}
K.~Hamaguchi, K.~Nakayama, and N.~Yokozaki, {\it {NMSSM in gauge-mediated SUSY
  breaking without domain wall problem}},  {\em Phys. Lett.} {\bf B708} (2012)
  100--106, [\href{http://arxiv.org/abs/1107.4760}{{\tt arXiv:1107.4760}}].

\bibitem{Kadota:2015dza}
K.~Kadota, M.~Kawasaki, and K.~Saikawa, {\it {Gravitational waves from domain
  walls in the next-to-minimal supersymmetric standard model}},  {\em JCAP}
  {\bf 1510} (2015), no.~10 041, [\href{http://arxiv.org/abs/1503.06998}{{\tt
  arXiv:1503.06998}}].

\bibitem{Hattori:2015xla}
H.~Hattori, T.~Kobayashi, N.~Omoto, and O.~Seto, {\it {Entropy production by
  domain wall decay in the NMSSM}},  {\em Phys. Rev.} {\bf D92} (2015), no.~10
  103518, [\href{http://arxiv.org/abs/1510.03595}{{\tt arXiv:1510.03595}}].

\bibitem{Saikawa:2017hiv}
K.~Saikawa, {\it {A review of gravitational waves from cosmic domain walls}},
  {\em Universe} {\bf 3} (2017), no.~2 40,
  [\href{http://arxiv.org/abs/1703.02576}{{\tt arXiv:1703.02576}}].

\bibitem{Derendinger:1983bz}
J.~P. Derendinger and C.~A. Savoy, {\it {Quantum Effects and SU(2) x U(1)
  Breaking in Supergravity Gauge Theories}},  {\em Nucl. Phys.} {\bf B237}
  (1984) 307--328.

\bibitem{Stephan:1997ds}
A.~Stephan, {\it {Dark matter constraints in the minimal and nonminimal SUSY
  standard model}},  {\em Phys. Rev.} {\bf D58} (1998) 035011,
  [\href{http://arxiv.org/abs/hep-ph/9709262}{{\tt hep-ph/9709262}}].

\bibitem{Griest:1990kh}
K.~Griest and D.~Seckel, {\it {Three exceptions in the calculation of relic
  abundances}},  {\em Phys. Rev.} {\bf D43} (1991) 3191--3203.

\bibitem{Ellwanger:2004xm}
U.~Ellwanger, J.~F. Gunion, and C.~Hugonie, {\it {NMHDECAY: A Fortran code for
  the Higgs masses, couplings and decay widths in the NMSSM}},  {\em JHEP} {\bf
  02} (2005) 066, [\href{http://arxiv.org/abs/hep-ph/0406215}{{\tt
  hep-ph/0406215}}].

\bibitem{Ellwanger:2005dv}
U.~Ellwanger and C.~Hugonie, {\it {NMHDECAY 2.0: An Updated program for
  sparticle masses, Higgs masses, couplings and decay widths in the NMSSM}},
  {\em Comput. Phys. Commun.} {\bf 175} (2006) 290--303,
  [\href{http://arxiv.org/abs/hep-ph/0508022}{{\tt hep-ph/0508022}}].

\bibitem{Belanger:2005kh}
G.~Belanger, F.~Boudjema, C.~Hugonie, A.~Pukhov, and A.~Semenov, {\it {Relic
  density of dark matter in the NMSSM}},  {\em JCAP} {\bf 0509} (2005) 001,
  [\href{http://arxiv.org/abs/hep-ph/0505142}{{\tt hep-ph/0505142}}].

\bibitem{Ellis:2018dmb}
J.~Ellis, N.~Nagata, and K.~A. Olive, {\it {Uncertainties in WIMP Dark Matter
  Scattering Revisited}},  {\em Eur. Phys. J.} {\bf C78} (2018), no.~7 569,
  [\href{http://arxiv.org/abs/1805.09795}{{\tt arXiv:1805.09795}}].

\bibitem{Staub:2015aea}
F.~Staub, P.~Athron, U.~Ellwanger, R.~Gr{\"o}ber, M.~M{\"u}hlleitner,
  P.~Slavich, and A.~Voigt, {\it {Higgs mass predictions of public NMSSM
  spectrum generators}},  {\em Comput. Phys. Commun.} {\bf 202} (2016)
  113--130, [\href{http://arxiv.org/abs/1507.05093}{{\tt arXiv:1507.05093}}].

\bibitem{Hisano:2013cqa}
J.~Hisano, T.~Kuwahara, and N.~Nagata, {\it {Grand Unification in High-scale
  Supersymmetry}},  {\em Phys. Lett.} {\bf B723} (2013) 324--329,
  [\href{http://arxiv.org/abs/1304.0343}{{\tt arXiv:1304.0343}}].

\bibitem{ATLAS:2018yhd}
{\bf ATLAS} Collaboration, T.~A. collaboration, {\it {Search for supersymmetry
  in final states with missing transverse momentum and multiple $b$-jets in
  proton-proton collisions at $\sqrt{s} = 13$ TeV with the ATLAS detector}}, .

\bibitem{Aaboud:2017vwy}
{\bf ATLAS} Collaboration, M.~Aaboud et~al., {\it {Search for squarks and
  gluinos in final states with jets and missing transverse momentum using 36
  fb$^{-1}$ of $\sqrt{s}=13$ TeV pp collision data with the ATLAS detector}},
  {\em Phys. Rev.} {\bf D97} (2018), no.~11 112001,
  [\href{http://arxiv.org/abs/1712.02332}{{\tt arXiv:1712.02332}}].

\bibitem{Aaboud:2017hrg}
{\bf ATLAS} Collaboration, M.~Aaboud et~al., {\it {Search for supersymmetry in
  final states with missing transverse momentum and multiple $b$-jets in
  proton-proton collisions at $ \sqrt{s}=13 $ TeV with the ATLAS detector}},
  {\em JHEP} {\bf 06} (2018) 107, [\href{http://arxiv.org/abs/1711.01901}{{\tt
  arXiv:1711.01901}}].

\bibitem{Sirunyan:2018vjp}
{\bf CMS} Collaboration, A.~M. Sirunyan et~al., {\it {Search for natural and
  split supersymmetry in proton-proton collisions at $ \sqrt{s}=13 $ TeV in
  final states with jets and missing transverse momentum}},  {\em JHEP} {\bf
  05} (2018) 025, [\href{http://arxiv.org/abs/1802.02110}{{\tt
  arXiv:1802.02110}}].

\bibitem{Aprile:2018dbl}
{\bf XENON} Collaboration, E.~Aprile et~al., {\it {Dark Matter Search Results
  from a One Tonne$\times$Year Exposure of XENON1T}},
  \href{http://arxiv.org/abs/1805.12562}{{\tt arXiv:1805.12562}}.

\bibitem{Akerib:2018lyp}
{\bf LUX-ZEPLIN} Collaboration, D.~S. Akerib et~al., {\it {Projected WIMP
  Sensitivity of the LUX-ZEPLIN (LZ) Dark Matter Experiment}},
  \href{http://arxiv.org/abs/1802.06039}{{\tt arXiv:1802.06039}}.

\bibitem{Billard:2013qya}
J.~Billard, L.~Strigari, and E.~Figueroa-Feliciano, {\it {Implication of
  neutrino backgrounds on the reach of next generation dark matter direct
  detection experiments}},  {\em Phys.Rev.} {\bf D89} (2014) 023524,
  [\href{http://arxiv.org/abs/1307.5458}{{\tt arXiv:1307.5458}}].

\bibitem{Aprile:2015uzo}
{\bf XENON} Collaboration, E.~Aprile et~al., {\it {Physics reach of the XENON1T
  dark matter experiment}},  {\em JCAP} {\bf 1604} (2016), no.~04 027,
  [\href{http://arxiv.org/abs/1512.07501}{{\tt arXiv:1512.07501}}].

\bibitem{Cremmer:1983bf}
E.~Cremmer, S.~Ferrara, C.~Kounnas, and D.~V. Nanopoulos, {\it {Naturally
  Vanishing Cosmological Constant in N=1 Supergravity}},  {\em Phys. Lett.}
  {\bf 133B} (1983) 61.

\bibitem{Ellis:1983sf}
J.~R. Ellis, A.~B. Lahanas, D.~V. Nanopoulos, and K.~Tamvakis, {\it {No-Scale
  Supersymmetric Standard Model}},  {\em Phys. Lett.} {\bf 134B} (1984) 429.

\bibitem{Lahanas:1986uc}
A.~B. Lahanas and D.~V. Nanopoulos, {\it {The Road to No Scale Supergravity}},
  {\em Phys. Rept.} {\bf 145} (1987) 1.

\bibitem{Kaplan:1999ac}
D.~E. Kaplan, G.~D. Kribs, and M.~Schmaltz, {\it {Supersymmetry breaking
  through transparent extra dimensions}},  {\em Phys. Rev.} {\bf D62} (2000)
  035010, [\href{http://arxiv.org/abs/hep-ph/9911293}{{\tt hep-ph/9911293}}].

\bibitem{Chacko:1999mi}
Z.~Chacko, M.~A. Luty, A.~E. Nelson, and E.~Ponton, {\it {Gaugino mediated
  supersymmetry breaking}},  {\em JHEP} {\bf 01} (2000) 003,
  [\href{http://arxiv.org/abs/hep-ph/9911323}{{\tt hep-ph/9911323}}].

\bibitem{Ellis:2001kg}
J.~R. Ellis, D.~V. Nanopoulos, and K.~A. Olive, {\it {Lower limits on soft
  supersymmetry breaking scalar masses}},  {\em Phys. Lett.} {\bf B525} (2002)
  308--314, [\href{http://arxiv.org/abs/hep-ph/0109288}{{\tt hep-ph/0109288}}].

\bibitem{Ellis:2010ip}
J.~Ellis, A.~Mustafayev, and K.~A. Olive, {\it {What if Supersymmetry Breaking
  Unifies beyond the GUT Scale?}},  {\em Eur. Phys. J.} {\bf C69} (2010)
  201--217, [\href{http://arxiv.org/abs/1003.3677}{{\tt arXiv:1003.3677}}].

\bibitem{Ellis:2016tjc}
J.~Ellis, J.~L. Evans, A.~Mustafayev, N.~Nagata, and K.~A. Olive, {\it {The
  Super-GUT CMSSM Revisited}},  {\em Eur. Phys. J.} {\bf C76} (2016), no.~11
  592, [\href{http://arxiv.org/abs/1608.05370}{{\tt arXiv:1608.05370}}].

\bibitem{Schmaltz:2000gy}
M.~Schmaltz and W.~Skiba, {\it {Minimal gaugino mediation}},  {\em Phys. Rev.}
  {\bf D62} (2000) 095005, [\href{http://arxiv.org/abs/hep-ph/0001172}{{\tt
  hep-ph/0001172}}].

\bibitem{Schmaltz:2000ei}
M.~Schmaltz and W.~Skiba, {\it {The Superpartner spectrum of gaugino
  mediation}},  {\em Phys. Rev.} {\bf D62} (2000) 095004,
  [\href{http://arxiv.org/abs/hep-ph/0004210}{{\tt hep-ph/0004210}}].

\bibitem{Ellis:2017djk}
J.~Ellis, J.~L. Evans, N.~Nagata, D.~V. Nanopoulos, and K.~A. Olive, {\it
  {No-Scale SU(5) Super-GUTs}},  {\em Eur. Phys. J.} {\bf C77} (2017), no.~4
  232, [\href{http://arxiv.org/abs/1702.00379}{{\tt arXiv:1702.00379}}].

\bibitem{Ibanez:1982fr}
L.~E. Ibanez and G.~G. Ross, {\it {SU(2)-L x U(1) Symmetry Breaking as a
  Radiative Effect of Supersymmetry Breaking in Guts}},  {\em Phys. Lett.} {\bf
  110B} (1982) 215--220.

\bibitem{Inoue:1982pi}
K.~Inoue, A.~Kakuto, H.~Komatsu, and S.~Takeshita, {\it {Aspects of Grand
  Unified Models with Softly Broken Supersymmetry}},  {\em Prog. Theor. Phys.}
  {\bf 68} (1982) 927. [Erratum: Prog. Theor. Phys.70,330(1983)].

\bibitem{Ibanez:1982ee}
L.~E. Ibanez, {\it {Locally Supersymmetric SU(5) Grand Unification}},  {\em
  Phys. Lett.} {\bf 118B} (1982) 73--78.

\bibitem{Ellis:1982wr}
J.~R. Ellis, D.~V. Nanopoulos, and K.~Tamvakis, {\it {Grand Unification in
  Simple Supergravity}},  {\em Phys. Lett.} {\bf 121B} (1983) 123--129.

\bibitem{Ellis:1983bp}
J.~R. Ellis, J.~S. Hagelin, D.~V. Nanopoulos, and K.~Tamvakis, {\it {Weak
  Symmetry Breaking by Radiative Corrections in Broken Supergravity}},  {\em
  Phys. Lett.} {\bf 125B} (1983) 275.

\bibitem{AlvarezGaume:1983gj}
L.~Alvarez-Gaume, J.~Polchinski, and M.~B. Wise, {\it {Minimal Low-Energy
  Supergravity}},  {\em Nucl. Phys.} {\bf B221} (1983) 495.

\bibitem{Matsuda:1995ta}
M.~Matsuda and M.~Tanimoto, {\it {Explicit CP violation of the Higgs sector in
  the next-to-minimal supersymmetric standard model}},  {\em Phys. Rev.} {\bf
  D52} (1995) 3100--3107, [\href{http://arxiv.org/abs/hep-ph/9504260}{{\tt
  hep-ph/9504260}}].

\bibitem{Haba:1996bg}
N.~Haba, {\it {Explicit CP violation in the Higgs sector of the next-to-minimal
  supersymmetric standard model}},  {\em Prog. Theor. Phys.} {\bf 97} (1997)
  301--310, [\href{http://arxiv.org/abs/hep-ph/9608357}{{\tt hep-ph/9608357}}].

\bibitem{Ham:2001kf}
S.~W. Ham, J.~Kim, S.~K. Oh, and D.~Son, {\it {The Charged Higgs boson in the
  next-to-minimal supersymmetric standard model with explicit CP violation}},
  {\em Phys. Rev.} {\bf D64} (2001) 035007,
  [\href{http://arxiv.org/abs/hep-ph/0104144}{{\tt hep-ph/0104144}}].

\bibitem{Garisto:1993ms}
R.~Garisto, {\it {Moderate supersymmetric CP violation}},  {\em Phys. Rev.}
  {\bf D49} (1994) 4820--4825, [\href{http://arxiv.org/abs/hep-ph/9311249}{{\tt
  hep-ph/9311249}}].

\bibitem{Ham:2001wt}
S.~W. Ham, S.~K. Oh, and D.~Son, {\it {Neutral Higgs sector of the
  next-to-minimal supersymmetric standard model with explicit CP violation}},
  {\em Phys. Rev.} {\bf D65} (2002) 075004,
  [\href{http://arxiv.org/abs/hep-ph/0110052}{{\tt hep-ph/0110052}}].

\bibitem{Ham:2003jf}
S.~W. Ham, Y.~S. Jeong, and S.~K. Oh, {\it {Radiative CP violation in the Higgs
  sector of the next-to-minimal supersymmetric model}},
  \href{http://arxiv.org/abs/hep-ph/0308264}{{\tt hep-ph/0308264}}.

\bibitem{Kuzmin:1985mm}
V.~A. Kuzmin, V.~A. Rubakov, and M.~E. Shaposhnikov, {\it {On the Anomalous
  Electroweak Baryon Number Nonconservation in the Early Universe}},  {\em
  Phys. Lett.} {\bf 155B} (1985) 36.

\bibitem{Pietroni:1992in}
M.~Pietroni, {\it {The Electroweak phase transition in a nonminimal
  supersymmetric model}},  {\em Nucl. Phys.} {\bf B402} (1993) 27--45,
  [\href{http://arxiv.org/abs/hep-ph/9207227}{{\tt hep-ph/9207227}}].

\bibitem{Davies:1996qn}
A.~T. Davies, C.~D. Froggatt, and R.~G. Moorhouse, {\it {Electroweak
  baryogenesis in the next-to-minimal supersymmetric model}},  {\em Phys.
  Lett.} {\bf B372} (1996) 88--94,
  [\href{http://arxiv.org/abs/hep-ph/9603388}{{\tt hep-ph/9603388}}].

\bibitem{Huber:2000mg}
S.~J. Huber and M.~G. Schmidt, {\it {Electroweak baryogenesis: Concrete in a
  SUSY model with a gauge singlet}},  {\em Nucl. Phys.} {\bf B606} (2001)
  183--230, [\href{http://arxiv.org/abs/hep-ph/0003122}{{\tt hep-ph/0003122}}].

\bibitem{Carena:2011jy}
M.~Carena, N.~R. Shah, and C.~E.~M. Wagner, {\it {Light Dark Matter and the
  Electroweak Phase Transition in the NMSSM}},  {\em Phys. Rev.} {\bf D85}
  (2012) 036003, [\href{http://arxiv.org/abs/1110.4378}{{\tt
  arXiv:1110.4378}}].

\bibitem{Cheung:2012pg}
K.~Cheung, T.-J. Hou, J.~S. Lee, and E.~Senaha, {\it {Singlino-driven
  Electroweak Baryogenesis in the Next-to-MSSM}},  {\em Phys. Lett.} {\bf B710}
  (2012) 188--191, [\href{http://arxiv.org/abs/1201.3781}{{\tt
  arXiv:1201.3781}}].

\bibitem{Demidov:2016wcv}
S.~V. Demidov, D.~S. Gorbunov, and D.~V. Kirpichnikov, {\it {Split NMSSM with
  electroweak baryogenesis}},  {\em JHEP} {\bf 11} (2016) 148,
  [\href{http://arxiv.org/abs/1608.01985}{{\tt arXiv:1608.01985}}]. [Erratum:
  JHEP08,080(2017)].

\bibitem{Akula:2017yfr}
S.~Akula, C.~Balázs, L.~Dunn, and G.~White, {\it {Electroweak baryogenesis in
  the $ {\mathbb{Z}}_3 $ -invariant NMSSM}},  {\em JHEP} {\bf 11} (2017) 051,
  [\href{http://arxiv.org/abs/1706.09898}{{\tt arXiv:1706.09898}}].

\bibitem{Bian:2017wfv}
L.~Bian, H.-K. Guo, and J.~Shu, {\it {Gravitational Waves, baryon asymmetry of
  the universe and electric dipole moment in the CP-violating NMSSM}},  {\em
  Chin. Phys.} {\bf C42} (2018), no.~9 093106,
  [\href{http://arxiv.org/abs/1704.02488}{{\tt arXiv:1704.02488}}].

\bibitem{King:2015oxa}
S.~F. King, M.~Muhlleitner, R.~Nevzorov, and K.~Walz, {\it {Exploring the
  CP-violating NMSSM: EDM Constraints and Phenomenology}},  {\em Nucl. Phys.}
  {\bf B901} (2015) 526--555, [\href{http://arxiv.org/abs/1508.03255}{{\tt
  arXiv:1508.03255}}].

\bibitem{Barr:1990vd}
S.~M. Barr and A.~Zee, {\it {Electric Dipole Moment of the Electron and of the
  Neutron}},  {\em Phys. Rev. Lett.} {\bf 65} (1990) 21--24. [Erratum: Phys.
  Rev. Lett.65,2920(1990)].

\bibitem{Nagata:2014wma}
N.~Nagata and S.~Shirai, {\it {Higgsino Dark Matter in High-Scale
  Supersymmetry}},  {\em JHEP} {\bf 01} (2015) 029,
  [\href{http://arxiv.org/abs/1410.4549}{{\tt arXiv:1410.4549}}].

\bibitem{Fukuda:2017jmk}
H.~Fukuda, N.~Nagata, H.~Otono, and S.~Shirai, {\it {Higgsino Dark Matter or
  Not: Role of Disappearing Track Searches at the LHC and Future Colliders}},
  {\em Phys. Lett.} {\bf B781} (2018) 306--311,
  [\href{http://arxiv.org/abs/1703.09675}{{\tt arXiv:1703.09675}}].

\bibitem{Baron:2013eja}
{\bf ACME} Collaboration, J.~Baron et~al., {\it {Order of Magnitude Smaller
  Limit on the Electric Dipole Moment of the Electron}},  {\em Science} {\bf
  343} (2014) 269--272, [\href{http://arxiv.org/abs/1310.7534}{{\tt
  arXiv:1310.7534}}].

\bibitem{2012NJPh14j3051K}
D.~M. {Kara}, I.~J. {Smallman}, J.~J. {Hudson}, B.~E. {Sauer}, M.~R. {Tarbutt},
  and E.~A. {Hinds}, {\it {Measurement of the electron's electric dipole moment
  using YbF molecules: methods and data analysis}},  {\em New Journal of
  Physics} {\bf 14} (Oct., 2012) 103051,
  [\href{http://arxiv.org/abs/1208.4507}{{\tt arXiv:1208.4507}}].

\bibitem{Kawall:2011zz}
D.~Kawall, {\it {Searching for the electron EDM in a storage ring}},  {\em J.
  Phys. Conf. Ser.} {\bf 295} (2011) 012031.

\bibitem{KKN}
J.~Kawamura, T.~Kobayashi, and N.~Nagata.
\newblock in preparation.

\end{thebibliography}\endgroup
}

\end{document}